\documentclass[nohyper,notoc]{JHEP}

\usepackage{amsmath,euscript,array,amssymb,cite} 
\newcommand{\startappendix}{
\setcounter{section}{0}
\renewcommand{\thesection}{\Alph{section}}}

\newcommand{\Appendix}[1]{
\refstepcounter{section}
\begin{flushleft}
{\large\bf Appendix \thesection: #1}
\end{flushleft}}

\def\aD{{\dot\alpha}}
\def\bD{{\dot\beta}}

\def\N{{\cal N}}

\def\sst{\scriptscriptstyle}

\newcommand{\BN}{\boldsymbol{N}}
\newcommand{\Bk}{\boldsymbol{k}}

\def\Dbarslash{\,\,{\raise.15ex\hbox{/}\mkern-12mu {\bar\D}}}
\def\Dslash{\,\,{\raise.15ex\hbox{/}\mkern-12mu \D}}
\def\delslash{\,\,{\raise.15ex\hbox{/}\mkern-9mu \partial}}
\def\delbarslash{\,\,{\raise.15ex\hbox{/}\mkern-9mu {\bar\partial}}}

\def\ms{{\mathfrak M}}

\def\Z{{\EuScript Z}}
\def\I{{\EuScript I}}

\def\B{{\EuScript B}}
\def\H{{\EuScript H}}

\newcommand{\EQ}[1]{\begin{equation} #1 \end{equation}}
\newcommand{\AL}[1]{\begin{subequations}\begin{align} #1
\end{align}\end{subequations}}
\newcommand{\SP}[1]{\begin{equation}\begin{split} #1 \end{split}\end{equation}}

\title{Notes on Soliton Bound-State Problems in Gauge Theory
and String Theory}

\author{Nick~Dorey$^{a}$, Timothy J.~Hollowood$^{a}$
and Valentin V.~Khoze$^b$\\
$^a$Department of Physics, University of Wales Swansea,
Swansea, SA2 8PP, UK\\
$^b$Department of Physics and IPPP, University of Durham,
Durham, DH1 3LE, UK\\
E-mail: {\tt n.dorey@swan.ac.uk}, {\tt t.hollowood@swan.ac.uk},
{\tt valya.khoze@durham.ac.uk}}

\abstract{We review four basic examples
where string theory and/or field theory
dualities predict the existence of soliton bound-states.
These include the existence of threshold bound-states of D0 branes
required by IIA/M duality and the closely-related bound-states of instantons
in the maximally supersymmetric five dimensional gauge theory. In the IIB
theory we discuss $(p,q)$-strings as bound-states of $D$ and $F$ strings,
as well as the corresponding bound-states of monopoles and dyons
in ${\cal N}=4$ supersymmetric Yang-Mills theory whose existence was
predicted by Sen. In particular we consider the
${\EuScript L}^2$-index theory relevant for counting these states.
In each case we show that the bulk contribution to the index can be evaluated
by relating it to an instanton effect in the corresponding theory with a
compact Euclidean time dimension. The boundary contribution to the index
can be determined
by considering the asymptotic regions of the relevant moduli space.}


\preprint{{\tt hep-th/0105090}\\SWAT-\\IPPP/01/20}

\begin{document}

\section{Introduction}

Dualities in supersymmetric field theory and string theory, which relate
strong and weak coupling, generally provide few predictions
which can be tested against existing knowledge. However, theories with
extended supersymmetry typically have BPS states
which form short representations of the supersymmetry
algebra. Consequently, in favourable conditions, the
BPS mass spectrum is stable under variations of the
coupling constant and the existence of BPS states required by duality
can often be tested directly at weak coupling.

Two key examples of this phenomenon arise in Type II string theory on flat
ten-dimensional space.
The first example arises in the IIA theory. Here, the duality
between IIA string theory and M-theory compactified on a circle
\cite{Witvarious},
requires the presence of an infinite tower of BPS states corresponding
to Kaluza-Kein modes of the eleven-dimensional metric. These states
are realized as threshold bound states of an arbitrary number of
D0-branes. The second example occurs in IIB string theory.
The IIB theory is believed to have an exact $SL(2,{\mathbb Z})$ duality,
known as S-duality, which acts by modular transformations on the
complexified IIB coupling $\tau_{\rm IIB}$.
The action of S-duality on the fundamental string leads
to a prediction for the existence of an $SL(2,{\mathbb Z})$ multiplet of
BPS saturated strings \cite{Schwarz}.
These states can be thought of as bound-states of
$q$ fundamental strings with $k$ D-strings. The
$SL(2,{\mathbb Z})$ orbit of the fundamental string contains states
corresponding to all coprime pairs $q$ and $k$. When $q$ and
$k$ have a common factor $l$, a $(q,k)$-string would be
exactly at threshold for decay into $l$, $(q/l,k/l)$ strings.
In fact various arguments suggest that there are no normalizable
bound-states with non-coprime values of the charges.
In the following,
we will call these two bound state problems the ``IIA'' and ``IIB''
examples, respectively.

Both these examples have close analogs in supersymmetric gauge
theories with sixteen supercharges. Most famously,
Montonen-Olive duality \cite{OM} of ${\cal N}=4$ SUSY Yang-Mills theory
with gauge group $SU(2)$ (or $U(2)$), predicts an infinite spectrum of BPS
saturated dyons with coprime integral electric and magnetic charges, $q$ and
$k$. As in the IIB example, bound-states with non-coprime values
of $q$ and $k$, which would be exactly at threshold, do not occur. The
connection to IIB string theory is that the ${\cal N}=4$ theory with
gauge group $U(N)$ can be realized on the world-volume of $N$ D3-branes
in the IIB theory. In this context Montonen-Olive duality of the
$U(2)$ gauge theory is mapped onto S-duality of the IIB theory. The
$(q,k)$-dyons of the $U(2)$ theory are realized as segments of
$(q,k)$-strings stretching between the two D3-branes.

More recently a
similarly precise gauge theory analog of the problem of D0-brane
bound-states in the IIA theory has emerged. Self-dual configurations of
a non-abelian gauge field are most familiar as instantons of finite
action in four spacetime dimensions, but they also appear as topologically
stable solitons in five dimensional gauge theory. A recent conjecture
relates the strong coupling limit of the five dimensional gauge theory
with sixteen supercharges to a theory with chiral $\N=(2,0)$
supersymmetry in six dimensions \cite{Rozali,Berkooz}.
The appearance of the additional
sixth dimension is analogous to the appearance of the eleventh
direction in M-theory. The Kaluza-Klein modes of the six-dimensional
fields correspond to threshold bound-states of Yang-Mills instantons
thought of a solitons in five dimensions. Again, the
connection to string theory emerges naturally when we realize the
gauge theory on the world volume of a D-brane. In this case, the
five dimensional theory with gauge group $U(N)$ appears on the world
volume of $N$ D4-branes. In this context, Yang-Mills instantons appear
as D0-branes which live on the D4-brane world-volume. The required
instanton bound-states then correspond directly to the threshold
bound-states of D0-branes in the IIA theory.

Hence, the two gauge theory examples in five and four dimensions,
respectively,
are realized on the world volumes of additional D-branes which act
as ``spectators'' to the basic the IIA/B bound-state problems
described above. In the following, we will use this
perspective extensively. To avoid confusion we will refer to
the five- and four-dimensional gauge theory examples, and their
realizations in Type II string theory, as the
$\text{IIA}'$ and $\text{IIB}'$ examples, respectively.
In all of these examples, duality
predicts the existence of BPS saturated bound-states of
solitonic extended objects.\footnote{In the following, we will use the
word ``soliton'' loosely, to describe all of these objects.}
The natural way to study this phenomenon at weak coupling
is to analyse the effective theory which lives on the world-volume
of the soliton. After factoring our the center-of-mass
(COM) degrees-of-freedom, the required multi-soliton
bound-states correspond to normalizable supersymmetric ground states of
this world-volume theory.
In the IIA/B examples, we have $k$ D-branes in a ten-dimensional
spacetime and the world-volume
theory is a $U(k)$ supersymmetric gauge theory with sixteen
supercharges. In the $\text{IIA}'$/$\text{B}'$ cases, the
higher dimensional D-branes introduce matter in the fundamental
representation of $U(k)$, breaking half the supersymmetry of the
world-volume theory of the lower dimensional D-branes.

For gauge theory solitons,
the most familiar formulation of the
effective world-volume theory is as a quantum mechanical $\sigma$-model
with the soliton moduli space as the target space. This is essentially
a supersymmetric version of Manton's
moduli space approximation \cite{manton} for soliton dynamics.
On the other hand,
the fact that the gauge theory solitons of the
$\text{IIA}'$/$\text{B}'$
examples can be realized as D-branes means that
the corresponding world-volume theories should also have a description
as a supersymmetric gauge theory. The connection between the two
points-of-view is now well established: for $k$ gauge theory solitons, the
standard rules of D-brane calculus yield a description of the
world-volume dynamics as a quantum mechanical $U(k)$ gauge
theory with eight supercharges. In the $\text{IIA}'$ example, the
relevant $D$-term equations reproduce the non-linear
constraints of the ADHM construction and the ADHM moduli space of
Yang-Mills instantons emerges as the Higgs branch of the world-volume
gauge theory \cite{Witsmall,Douglas1,Douglas2}.
In the $\text{IIB}'$ example, the Nahm construction of
moduli space of $k$ BPS monopoles emerges in an analogous way
\cite{Diacon} as the
Higgs branch of an impurity theory \cite{Impure}. In both
cases, on the Higgs branch,
the $U(k)$ gauge coupling is irrelevant and the world-volume gauge
theory flows in the IR to a non-linear $\sigma$-model with the
Higgs branch as the target space. In other words, the conventional
moduli space description of the solitons is recovered at low energy.
This is consistent with the preservation of exactly eight supercharges,
because the moduli spaces of gauge theory monopoles and instantons are
non-trivial hyper-K\"{a}hler manifolds. In contrast,
for the basic IIA/B examples, the
preservation of sixteen supercharges requires the moduli space metric
to be flat.

In each of the cases described above, the basic problem is to find
normalizable supersymmetric ground-states of the world-volume theory
which describes the relative motion of two or more solitons.
A standard strategy to prove the existence of supersymmetric
ground-states is to compute the Witten index \cite{Windex},
$\I={\rm Tr}\,(-1)^{F}e^{-\beta\H}$.
If the index is equal to a positive
integer $l$, then the existence of at least $l$
supersymmetric ground-state is guarenteed. Note, however, that
the index does not
prevent the existence additional vacua which come in bose/fermi
pairs. As with other quantum mechanical systems in non-compact
spaces the problem of calculating the index is
complicated by the fact that $\H$ has a continuous spectrum of
scattering states. In cases involving soliton bound-states at threshold, the
problem is even more severe as there is no mass gap between the
required SUSY ground state and the continuum. These problems
neccesitate carefully defining the so-called
${\EuScript L}^2$-index of $\H$ which correctly counts normalizable
ground-states. As we review below, the ${\EuScript L}^2$-index
$\I_{{\EuScript L}^2}$ is the sum of a bulk contribution $\I_{\rm
bulk}$ which is the integral of an index density over the whole
configuration space and a ``defect'' term $\I_{\rm boundary}$
which comes from an integral over the boundary at infinity
\cite{Sethstern,Yi}.

One of the main common features of all the bound-state problems
described above is that the index can be evaluated by
exploiting the general connection between solitons in $d$-dimensional
spacetime
and instantons in $d-1$ dimensions. In string theory, this
connection is just a special case of a more general phenomena:
a finite action instanton can be obtained by wrapping the world-volume
of a $p$-brane around a
non-contractable $(p+1)$-dimensional cycle in a spacetime of Euclidean
signature \cite{Becker}. The wrapped object can be any BPS brane or
BPS bound-state of such branes. Multiple wrappings of the
brane world-volume must also be included with appropriate weights.
If we compactify the basic IIA example
on ${\mathbb R}^{9}\times S^{1}$, then each threshold bound-state of
D0-branes has a Euclidean worldline which can wind around the
compact direction an arbitrary number of times. After T-duality this
yields a series of D-instanton corrections to the ${\cal R}^{4}$ term
in the IIB effective action. Via this connection Green and Gutperle
\cite{GG1} showed that the coefficient of the $k$ D-instanton
contribution is essentially equal to the bulk contribution to the
${\EuScript L}^2$-index for the binding of $k$ D0-branes. However, the
${\cal  R}^{4}$ term is also highly constrained by supersymmetry and
by the S-duality of the IIB theory \cite{Sethgreen} and can be
determined exactly. The weak coupling expansion of the exact result can
then be used to derive the bulk contribution to the index for each
value of $k$.

A wide variety of exact
results for string theory effective actions can be understood in
this way (for a review see \cite{Kiritsis}). In particular, a simple
set of rules for D-brane instanton calculus have been deduced from the
constraints of U-duality \cite{PK}. One feature of these
rules is that
the contributions of multi-D-brane bound states appear on an equal footing
with those of single D-branes. In this respect D-brane bound-states are
``pointlike'', showing no indications of substructure.
The same rules can also be applied to
the $\text{IIA}'$/$\text{B}'$ examples where the D-branes have an
alternative description in terms of gauge theory solitons and
instantons. One of the aims of the present work
(and of \cite{part1,part2}) is to investigate how these rules emerge
from a conventional semiclassical analysis of the corresponding
gauge theory. In particular, we will find that gauge theory
instanton calculations reproduce characteristically D-braney features
such as the contribution of pointlike bound-states and of multiple
winding sectors.

In gauge theory, there is already a familiar relation between
solitons and instantons.
We will start from
a classical, static, BPS saturated $k$-soliton configuration with
finite mass $kM$ in a $d$-dimensional Lorentzian gauge theory.
We then Wick rotate and compactify
the resulting Euclidean time dimension on a circle of circumference
$\beta$ with SUSY preserving
boundary conditions. Now the static soliton
becomes an instanton of finite Euclidean action $\beta kM$ in the
compactified theory. Because of the BPS property, the resulting
instanton will be invariant under half the supersymmetry generators.
In a supersymmetric theory with $4{\cal N}$ supercharges
the instanton will then have $2{\cal N}$ exact
fermion zero modes\footnote{These are the so-called supersymmetric zero
modes. If all the vacuum expectation values are turned off an instanton
can have additional $2{\cal N}$ superconformal fermion zero modes.
} and will contribute to a correlation function with
the same number of fermionic insertions.
The key point is that each correlation function of the
compactifed theory can be interpreted as a
trace over the BPS sector of the Hilbert space of the original
theory \cite{part2}:
\begin{equation}
\langle \psi(x_{1})\psi(x_{2})\ldots \psi(x_{2{\cal N}}) \rangle
 = {\rm
Tr}_{\rm BPS}\left[\psi(x_{1})\psi(x_{2})\ldots \psi(x_{2{\cal N}})
(-1)^{F}\,e^{-\beta\H}\right]\ .
\label{trace}
\end{equation}
The correlators can therefore be thought of as
refinements of the Witten index.
Just as in favourable circumstances, the Witten index only gets
non-zero contributions from zero energy states,
so these correlators only get
contributions from BPS states. Similar phenomena in
two dimensional QFT have previously been discussed in \cite{CV}.

A BPS saturated $k$ soliton
bound-state of mass $kM$, contributes
a term of order $\exp(-\beta kM)$ to the trace on the right-hand side
of this equality. In the path integral language, this term can be
identified with the contribution of the corresponding
instanton in the compactified theory. In the semiclassical limit, we
can express the overall coefficient of the contribution
as an integral over the soliton
moduli space. The main result, which applies equally in the string and
gauge theory examples, is that the resulting integral is essentially
equal to the bulk contribution to the ${\EuScript L}^2$-index of the
corresponding bound-state problem in the original $d$-dimensional
theory. On the other hand the $2{\cal N}$-fermion correlator appearing
in \eqref{trace} can also be related to a local vertex of the
form $\bar{\psi}^{2{\cal N}}$ in the effective action of the
compactified theory. As in the IIA example, the
low-dimensional terms in this effective action are
highly constrained by supersymmetry can often be determined exactly.
In all the examples we will consider, obtaining the exact answer
requires some additional physical input from duality.
Finally, the coefficients of
instanton expansion of the exact result then provide a definite
prediction for bulk contribution to the ${\EuScript L}^2$-index which counts
multi-soliton bound-states.

The main goal of these notes is to show how a unified picture of
${\EuScript L}^2$-index theory relevant to each of these $k$ soliton bound-state
problems emerges. Strictly speaking, we will make only a little
progress towards a first principles calculation of the BPS spectrum.
However, numerous consistency checks between seemingly different
mainifestations of duality
in string theory and gauge theory will appear. Before describing the
details we list the main results:

{\bf 1:} In each case, the bulk contribution to the ${\EuScript L}^2$-index can be
expressed as the partition function of a $U(k)$ matrix model with
certain COM degrees-of-freedom factored off.
In the $\text{IIA}'$/$\text{B}'$ problems
the bulk contribution can also be represented as the
integral of an Euler density over a smooth
moduli space (after a suitable resolution of singularities in the
$\text{IIA}'$ case).

{\bf 2:} In each case, we find that the bulk contribution can
be calculated indirectly by relating it to a corresponding
instanton contribution to the Wilsonian effective action of a
compactified theory. We review the exact results for
instanton contributions which yield predictions for the bulk
contributions to the corresponding bound-state problem. In each case
the bulk contribution has the interpretation as a sum over wrapped
brane world-volumes. The bulk contribution overcounts the
${\EuScript L}^2$-index precisely by including the contribution of sectors
of multiple D-brane wrapping.

{\bf 3:} In the gauge theory examples, we can relate the bulk
contribution to the ${\EuScript L}^2$-index to the geometric Euler characteristic
of the moduli space using classical index theory for a manifold
with boundary. We find agreement with recent results on the homology
of instanton and
monopole moduli spaces due to Nakajima \cite{Nak} and to Segal and
Selby \cite{Segal}, respectively. Our results can be checked very
explicitly in the two soliton sector.

{\bf 4:} The defect term is only sensitive to the
asymptotic region of the moduli space where it can be calculated
without knowing the details of the interactions between solitons.
We will review and expand on a heuristic argument due to Yi \cite{Yi},
which is
effective in determining the defect term in each case. In each case
the total ${\EuScript L}^2$-index is consistent with the predictions of duality
discussed above.

In the remainder of this section, we discuss some general features of the
problems in question. In \S2 we review most of the elements
described above in the context of the best understood
string theory example: the problem of counting
threshold bound-states of D0-branes in IIA string theory on
${\mathbb R}^{9,1}$.
The discussion is then extended to the other cases in the remaining
sections.

In each of the examples we
will consider the world-volume dynamics of
$k$ D-branes. Although the dimension of the D-branes in question
will vary the common feature is that, in each case, all the spacelike
dimensions of the world-volume will be compact. Initially we will work
in a spacetime of Lorentzian signature and the single time-like
dimension on the world volume will be non-compact. Thus in all cases
the effective world-volume dynamics at low energy will be described by
a quantum-mechanical $U(k)$ gauge theory with adjoint, and for the
II$\text A'$/$\text B'$ cases fundamental, matter. The
eigenvalues of the adjoint-valued scalar fields describe the
positions of the $k$ D-branes in their common transverse directions.
In the basic IIA/B examples there are no additional branes and the
corresponding world-volume theory has sixteen supercharges.
These gauge theories have a Coulomb branch of gauge
equivalent classical
vacua corresponding to the configuration space of the $k$
branes. Since there is fundamental matter for the II$\text A'$/$\text
B'$ examples, there are
Coulomb, Higgs and mixed branches arising in the gauge theory examples
with eight supercharges.

In all the cases we will consider, the classical flat
directions cannot be lifted by quantum corrections because of
supersymmetry.
In a four-dimensional gauge theory, the corresponding
interpretation is that a moduli space of gauge-inequivalent
vacua persists at the quantum level. However in quantum mechanics,
or in $(1+1)$-dimensional field theory, this
language is not really appropriate as the
ground-state wavefunctions spread out over the manifold of classical
vacua. Nevertheless it has been established that the distinction
between the Higgs and Coulomb phases persists once quantum corrections
are taken into account \cite{Witten1997,Berkooz1999,Aharony1999}.
Moreover, just as in higher dimensions, the flat directions are
associated with massless modes and we can write down a Wilsonian
effective action for these degrees-of-freedom and order the terms
according to the number of time derivatives. As in the higher
dimensional cases, supersymmetry naturally pairs these time
derivatives with the
world-volume fermions, order by order in this expansion.

The time derivatives of the scalar fields which parametrize the
classcial moduli space correspond
to the velocities, $v_i$, $i=1,\ldots, k$, of the
$k$ D-branes.
The effective action of the $U(k)$ gauge theory
then has a non-trivial expansion in powers of these velocities.
The first non-trivial term in the velocity expansion has the form
$\tfrac12g_{ij}v^{i}v^{j}$ where the tensor $g_{ij}$ defines a metric on the
manifold of classical vacua $\ms_{cl}$. The supersymmetric
completion of this term defines a quantum mechanical supersymmetric
non-linear $\sigma$ model with target space $\ms_{cl}$.
In the basic IIA/B cases, the effective theory has sixteen
supercharges which forces the metric to be flat. In the
$\text{IIA}'$/$\text{B}'$ examples, the effective theory
has only eight supercharges and a non-trivial hyper-K\"{a}hler metric
is allowed. In these cases, the off-diagonal terms in the metric describe
the leading velocity-dependent interactions between gauge-theory
solitons. The order $v^{2}$ approximation to the
effective world-volume action coincides with Manton's moduli-space
approximation for soliton dynamics \cite{manton}.

\section{The IIA Bound-State Problem}

In the IIA case discussed above, the motion of $k$ D0-branes
is described by a quantum mechanical
$U(k)$ gauge theory with nine adjoint-valued scalar fields and
sixteen adjoint-valued fermions. The eigenvalues of the adjoint
scalars describe the relative position of the D0-branes in their nine
transverse directions. Along these flat directions, non-zero values of the
scalar fields break the gauge group down to its Cartan subalgebra
and we have a manifold of
gauge-inequivalent classical vacua. This Coulomb branch
is just $\text{Sym}_{k}
({\mathbb R}^{9})=({\mathbb R}^9)^k/S^k$ where the symmetric
product in the numerator is simply the configuration space of $k$
identical particles moving in nine dimensional space. The
permutation group $S_{k}$, representing the interchange
symmetry of $k$ identical particles, corresponds to the
Weyl subgroup of the $U(k)$ gauge group. An overall factor of
${\mathbb R}^9$
in the numerator corresponds to the COM
degrees-of-freedom of the D0-brane configuration. After seperating out these
modes, the relative motion of the branes is described by the
corresponding theory with gauge group $SU(k)$. The main problem is to
show that this theory has a single normalizable supersymmetric
ground-state for each value of $k$.

We begin by investigating the low energy dynamics of the D0-branes
in the velocity expansion described above.
The first non-trivial terms in the velocity
expansion define a quantum mechanical non-linear $\sigma$-model
with target space
$\text{Sym}_{k}({\mathbb R}^9)/{\mathbb R}^9$. As we have sixteen
supercharges, quantum corrections to the classical flat metric on the
Coulomb branch are forbidden. As there are no off-diagonal terms
in the metric the interactions between D0-branes vanish
at order $v^{2}$. In fact the first non-trivial interactions between
D0-branes occur at order $v^{4}$. The supersymmetric completion of
these interactions involves terms with up to eight fermions.
Note that the target space has orbifold singularities at the
fixed points of the Weyl group where one or more D0-branes coincide.
Just as in higher dimensions,
a non-abelian subgroup of the gauge group is restored at these points
and the low energy description breaks down due to the presence of
additional massless fields.
In the D0-brane case, the presence of singularities mean that
the velocity expansion is only reliable for describing the scattering
of D0-branes at large impact parameter.
There is no reason to expect this expansion
to be adequate for determining the existence or otherwise of
D0-brane bound-states. Indeed, as interactions between D0-branes vanish
at order $v^{2}$, its obvious that we will not find any bound-states in a
na\"\i ve ``moduli space'' approximation. More generally,
one expects that the behaviour of the wavefunction near
the origin depends on the full non-abelian dynamics.

As mentioned above, the standard approach to determining the
presence of supersymmetric ground-states is to compute the Witten index
$\I={\rm Tr}\,(-1)^{F}e^{-\beta\H}$.
The utility of the Witten
index comes from the fact that, modulo some caveats to be
discussed below, it is invariant under all deformations of the theory
which preserve supersymmetry. To avoid the
possibility of vacua going to infinity in field space as the
parameters are varied, we should restrict our attention to
variations which do not alter the asymptotic behaviour of the
potential. Such deformations can
often be used to reach a regime in parameter space where the index can
be computed easily. In particular one may use the $\beta$-independence of
the index to take the limit $\beta\rightarrow 0$, in which
the quantum mechanical path integral reduces to an ordinary
integral. Generally speaking, this proceedure is extremely robust
whenever the spectrum of the theory is discrete or, at least,
has a finite mass gap.

As discussed above, the D0-bound-states in question
are exactly at threshold.  Correspondingly, the $SU(k)$ world-volume theory
has flat directions which are not
lifted by quantum effects and the D0-brane moduli space is
non-compact. The main problem introduced
by non-compactness is the fact that the theory has
a continuous spectrum of scattering states in addition
to the discrete bound state spectrum.
In particular, even scattering states of non-zero
energy can actually contribute to the Witten index.
Na\"\i vely states of
non-zero energy come in bose-fermi pairs which cancel in the
Witten index due to the insertion of $(-1)^{F}$ appearing in the
trace. However, although
supersymmetry demands that the range of the
continuous spectrum is the same for bosons and fermions,
it does not necessarily require the density of
these states to be equal. A difference between the densities
of bose and fermi scattering states of non-zero energy
can then lead to anomalous $\beta$ dependence of the index \cite{cal}.
This contribution of states of non-zero energy can be eliminated
by defining the desired index
as the $\beta\rightarrow \infty$ limit of $\I(\beta)$.
However, we can no longer use the $\beta\rightarrow 0$ limit to
compute the index.

In the present case,
the problems related to non-compactness persist even in the
$\beta\rightarrow \infty$ limit because there
is no mass gap between the required
supersymmetric ground-states and the continuum.
However, a ground-state corresponding to a
D-brane bound-state must have a normalizable wavefunction and
this distinguishes it from the continuum of scattering states
which are non-normalizable.
Hence the trace appearing in the
Witten index should be restricted to states in the
Hilbert space which have square-integrable wave funtions.
In this case the Witten index coincides with what is
known in functional analysis as the ``${\EuScript L}^2$-index'' of the
Hamiltonian. We will now review the approach to
computing the ${\EuScript L}^2$-index developed by Sethi and Stern
\cite{Sethstern} and by Yi \cite{Yi}.

A sensible definition of the ${\EuScript L}^2$-index requires us to
regulate the problem of non-compactness before taking the
$\beta\rightarrow\infty$ limit. This can be accomplished
by restricting the configuration space to the interior of a ball
$B_{R}$ of finite radius $R$. Typically the regulated index will
depend on $\beta$ via the dimensionless parameter $\kappa=\beta/R$.
In the case of $k$
D0-branes of the IIA theory, the configuration space is the
classical Coulomb branch of the $SU(k)$ gauge theory on the
world-volume, $\text{Sym}({\mathbb R}^9)/{\mathbb R}^9$ and the ball just the
subspace where the distance of any brane from the centre of mass
is less than $R$. The regulator $R$, should
only be taken to infinity {\em after\/} taking the
$\beta\rightarrow\infty$ limit. Note that this corresponds to the
limit $\kappa\rightarrow \infty$. In contrast, removing the regulator
at finite $\beta$ corresponds to the opposite limit $\kappa\rightarrow
0$. Thus we have
\EQ{
\I_{{\EuScript L}^2} = \lim_{R \rightarrow \infty} \,
\lim_{\beta\rightarrow \infty}
\ \int_{B_R} d^px\, {\rm Tr}\,(-1)^Fe^{-\beta\H}(x,x)\ .
\label{bulk}
}
Here the notation $(x,x)$ denotes that the Trace is evaluated
between position space eigenstates centered at the point $x$ in
$B_{R}$.

Sethi and Stern then showed that the resulting index can be
written as the sum of bulk and boundary contributions:
$\I_{{\EuScript L}^2}=\I_{\rm bulk}+\I_{\rm boundary}$
with
\AL{
\I_{\rm bulk} & =  \lim_{R \rightarrow \infty} \,
\lim_{\beta\rightarrow 0}
\ \int_{B_{R}} d^{p}x\, {\rm Tr}\,(-1)^{F}e^{-\beta\H}\ ,
\label{bulk2}\\
\I_{\rm boundary} & = \lim_{R \rightarrow \infty} \,
\lim_{\beta\rightarrow 0}
\ \int_{\partial B_{R}} d^{p-1}x \,
\int _{\beta}^{\infty}\, d\beta'\,
{\rm Tr}\,e_{n}(-1)^{F}Qe^{-\beta'\H}\ ,
\label{surface}
}
where $Q$ denotes the supercharge and $e_{n}$ is a fermion component in the
direction normal to the
boundary, $\partial B_{R}$, of $B_{R}$. For more details see Section 3 of
\cite{Sethstern}.

Following Green and Gutperle \cite{GG1}, the bulk contribution to the index
can be calculated by determining a corresponding instanton
contribution to the effective action. We now
compactify the Euclidean IIA theory down to nine dimensions
on a spacelike circle of circumference $\beta$. The D0-bound-states
of the ten-dimensional theory become instantons of finite action after
compactification. In particular, the relevant instanton
configurations involve D0-brane bound-state with worldlines
wrapped around the compact dimension. The BPS configurations are those
which involve a single worldline and are labelled by an integer
winding number $l$ as well as the number, $m$,
of constituent D0-branes. When
$\beta\ll\sqrt{\alpha'}$, it is appropriate
to perform T-duality to the IIB theory on the dual circle. Under
a T-duality transformation of the compact dimension,
the wrapped D0-brane worldlines of the IIA theory become D-instantons
of the IIB theory. These D-instantons have sixteen unlifted zero modes
and contribute at leading semiclassical order to a sixteen fermion
vertex which is contained in the supersymmetric completion of the
${\cal R}^{4}$ term in the IIB effective action. On the one hand, the
D-instanton contributions can be related to the bulk contribution to
the ${\EuScript L}^2$-index considered above. On the other hand, the corresponding
term in the IIB action can be determined exactly using
supersymmetry and the $SL(2,{\mathbb Z})$ invariance of the IIB theory
\cite{Sethgreen}. Briefly,
supersymmetry restricts that the coupling constant dependence to be
harmonic and S-duality means that we should therefore solve the
Laplace equation on the fundamental domain of $SL(2,{\mathbb Z})$.
As usual for harmonic functions, the desired solution is uniquely
determined by
specifying the boundary conditions
(as well as the absence of singularities). In the present case the
boundary conditions are determined by a tree level calculation at weak
coupling. Note that, in this approach, we are {\it assuming\/}
S-duality and the
final result which will be consistent with the existence of the
D0-brane bound-states required by IIA/M duality
should simply be regarded as a consistency check on the ``web of
dualities''.

Finally, Green and Gutperle obtained the explicit prediction,
\begin{equation}
\I_{\rm bulk}=\sum_{d|k} \frac{1}{d^{2}}\ .
\label{ibulk2}
\end{equation}
As the bulk piece corresponds to the
$\beta\rightarrow 0$ limit of the Witten index,
an explicit formula may be obtained by reducing the partition function
of the quantum mechanical gauge theory to zero dimensions.
For $k$ D0-branes, this yields an ordinary integral over
traceless $k\times k$ matrices modulo $SU(k)$ gauge
transformations. Initially this integral was evaluated explicitly for
the case of two D0-branes in \cite{Sethstern,Yi}, yielding
results in agreement with \eqref{ibulk2}.
Subsequently the integral was considered for all values of $k$,
by Moore, Nekrasov and Shatashvilli \cite{MNS},
again confirming \eqref{ibulk2} (see also \cite{KNS}).

The formula \eqref{ibulk2} has a nice interpretation in terms of the
original IIA theory on a circle. The term corresponding to the
divisor $d$, corresponds to the contribution of a bound-state of
$l=d$ D0-branes with world-line wrapped $m=k/d$ times round the
circle. This suggests (correctly) that the first term in the sum is the
contribution of the desired bound-state of $k$ D0-branes
wrapped only once around the compact dimension. What remains to be
explained is how the boundary contribution subtracts out the
spurious additional terms as well as the origin of the
weighting factor $1/d^{2}$ for each term in the sum.

The boundary contribution is hard to analyse in a precise way.
The complexity of the problem is due to the fact that the boundary
contains ``clustering regions'' where the seperation between one or
more subset of the $k$ D0-branes remains small.
This problem is absent
for $k=2$, and
first appears in the $k=3$ case where the
boundary contains a region where two D0-branes are far from the
third one but remain close to each other. Despite this problem,
there is a heuristic argument due to Yi \cite{Yi}
(and subsequently developed by
Green and Gutperle \cite{GG1}) which is effective in determining
the boundary contribution which we will now review. In the process we will
expand the original argument and place it in a more general
setting which will apply to the other bound-state problems we wish to
consider.

The basic argument is tantamount to the usual assumption in
quantum mechanics (and quantum field theory) that we can define
asymptotic scattering states. Essentially, this means that we
can split the Hamiltonian of the system as: $\H=
\H_{0}+\H_{\rm int}$.
Here $\H_{0}$ is an appropriate free Hamiltonian.
The asymptotic particle
states are eigenstates of $\H_{0}$ and $\H_{\rm int}$
accounts for the interactions between these states which are assumed
to have finite range. This is reasonable in the case of D0-branes as
the large distance
interactions between them are supressed by four powers of the velocity.
The velocity of states of fixed momentum is very small in the weak
coupling limit because their masses scale with an inverse power of the
coupling. In the first instance the asymptotic particle
states will be the D0-branes themselves, however we will
subsequently allow for the fact that bound-states of D0-branes
can also appear as asymptotic states using an inductive argument.

The idea is that the boundary
contribtion to the index comes from the asymptotic region where the
interactions between the D0-branes (and any other asymptotic states)
can be neglected. It can therefore
be calculated with respect to free Hamiltonian, $\H_{0}$. Thus,
we have
\begin{equation}
\I_{\rm boundary}= \I^{(0)}_{\rm boundary}=
\lim_{R \rightarrow \infty} \,
\lim_{\beta\rightarrow 0}\
\int_{\partial B_{R}} d^{p-1}x \,
\int _{\beta}^{\infty}\, d\beta'\,
{\rm Tr}\,e_{n}(-1)^{F}Qe^{-\beta'\H_{0}}\ .
\label{surface0}
\end{equation}
As the free Hamiltonian certainly has no bound-states we have,
$\I^{(0)}_{{\EuScript L}^2}= \I^{(0)}_{\rm bulk}+
\I^{(0)}_{\rm boundary}=0$. Consequently we have,
$\I_{\rm boundary}=
\I^{(0)}_{\rm boundary}=-\I^{(0)}_{\rm bulk}$
where,
\EQ{
\I^{0}_{\rm bulk}=\lim_{R \rightarrow \infty} \,
\lim_{\beta\rightarrow 0}\
\int_{B_{R}} d^{p}x\, {\rm Tr}\,(-1)^{F}e^{-\beta \H_{0}}\ .
\label{bulk0}
}

Neglecting the interactions D0-branes are simply massive
identical non-relativistic particles moving in ${\mathbb R}^9$.
Generalizing slightly, we will evaluate
\eqref{bulk0} in the case where, $\H_{0}$ is the
supersymmetric Hamiltonian of $k$ identical free massive
particles moving on
${\mathbb R}^m$ in the center of mass frame.
The bosonic degrees of freedom are therefore real variables
$X^i_a$ where $a=1,2,\ldots,b$ is a vector index of the
$SO(b)$ R-symmetry group and
$i=1,\ldots,k$ is an additional index labelling the $k$
particles. This system is supersymmetrized by
including $k$ fermions corresponding to the fermionic
zero modes of D0-branes. These Grassmann variables are denoted,
$\Psi^{i}_{\alpha}$ each with $f$
real components where $\alpha=1,2,\ldots,f$ is an appropriate
spinor index of the R-symmetry group.
In the case of D0-branes the relevant values of $b$ and $f$
are 9 and 16, respectively, but we will meet other cases below.

The relevant Hamiltonian is the sum of free Hamiltonians for
Bosons and Fermions: $\H^{0}=\H^{0}_{X}+
\H^{0}_{\Psi}$. However, we still
need to account for the interchange symmetry of the identical
particles. Specifically we must mod out by all permutations
$\pi\in S_{k}$ which act on the bose and fermi degrees of
freedom via the $k\times k$ representation matrix $M_{\pi}$:
\EQ{
\pi {\rm :} \qquad{} X^{a}_{i} \rightarrow
(M_{\pi})_{i}^{j}X_{j}^a\ ,
 \qquad
\Psi^{\alpha}_{i} \rightarrow (M_{\pi})_{i}^{j}\Psi^{\alpha}_{j}\ .
\label{pi}
}
To mod out the interchange symmetry we must insert a projector onto
$S_{k}$ invariant states inside a trace over the full Hilbert space.
\EQ{
\I^{0}_{\rm bulk} = \lim_{R \rightarrow \infty} \,
\lim_{\beta\rightarrow 0}\
\int_{{\mathbb R}^{9k}} d^{9k}x\, {\rm Tr}\,(-1)^{F}\,{\EuScript P}\,
e^{-\beta (\H^{0}_{X}+\H^{0}_{\Psi})}\ ,
\label{bulkp}
}
with
\begin{equation}
{\EuScript P}= \frac{1}{k!} \sum_{\pi \in S_{k}}\, M_{\pi}\ ,
\label{proj}
\end{equation}
where $M_{\pi}$ acts on the bose and fermi variables as in
\eqref{pi}. This sum over permutations has a very simple physical
interpretation in terms of particle worldlines. Recall that
an arbitrary permutation of $k$ objects admits a
decomposition into a product of mutually commuting
cyclic permutations of different lengths,
$l_{1},l_{2},\dots,l_{r}$ with $\sum_{j=1}^{r} l_{j}=k$.
A cyclic permutation of length one is trivial and
corresponds to the worldline of a single particle brane winding around the
compact dimension. The identity element of $S_{k}$ corresponds to a
product of cycles each of length one: $l_{j}=1$ for $j=1,\ldots k$.
This corresponds to $k$ particles each with a worldline which wraps
the compact dimension once. In contrast a
cyclic permutation of length
$l>1$ corresponds to a single particle with worldline wrapped around
the compact dimension $l$ times. A generic element of $S_{k}$
specified by integers $l_{j}$ with $\sum_{j=1}^{r} l_{j}=k$ above
corresponds to a configuration with $r$ particles where the
worldline of the $j^{\rm th}$ particle wraps the compact dimension $l_{j}$
times.

The traces \eqref{bulkp} are elementary to evaluate via a path
integral represtentation in the
$\beta\rightarrow 0$ limit and yield,
\begin{equation}
\I^{0}_{\rm bulk}=\frac{1}{k!} \sum_{\pi \in S_{k}}\,
\Big({\rm det}\left[1-M_{\pi}\right]\Big)^{\frac{1}{2}f-b}\ .
\label{proj2}
\end{equation}
As explained in \cite{GG1}, the only
non-vanishing contribution to the sum comes from the
$(k-1)!$ cyclic permutations of length $k$, for which
${\rm det}\left[1-M_{\pi}\right]=k$. Thus the states
with worldline wrapped $k$ times around the compact dimension
contribute to the partition function. The natural
interpretation is that multi-particle states have additional exact
fermion zero modes beyond those associated with the COM
degrees-of-freedom which we have already modded out. We obtain,
$\I^{0}_{\rm bulk}=k^{\frac{1}{2}f-b-1}$.
In the D0-brane case we have $\frac{1}{2}f-b=8-9=-1$ and thus we find that
$\I_{\rm boundary}=-\I^{0}_{\rm bulk}=-k^{-2}$.

We have so far accounted only for the contributions of the
D0-branes themselves. However, if bound-states of D0-branes exist it
is natural to expect that these also appear as asymptotic states. In
the semiclassical language, this corresponds to the asymptotic regions
of the moduli space in which one has several ``clusters'' of D0-branes with a
large distance between clusters.
Obviously, we must beware of making a circular
argument here as the presence of bound-states is precisely what we are
trying to demonstrate! However, bearing this in mind, we may make the
following inductive argument. Let us assume that a single threshold
bound-state of $k$ D0-branes is present in the spectrum for all $k<k'$.
We will now analyse the boundary contribution in the case $k=k'$.
By assumption, the possible asymptotic states
correspond to all possible sets of D0-brane bound-states with total D0
brane number $k=k'$. As before we treat the constituents as
non-interacting particles and look for the non-zero contributions to
the bulk index. By the same reasoning used above,
the only non-zero contributions come from
cyclic permutations of all the particles. Such a permutation only
occurs when all the particles present are identical. Thus the only
non-zero contributions come from sectors with $d>1$ identical
bound-states, each with D0-number $k'/d$. Thus, for $k=k'$, we have,
\begin{equation}
\I_{\rm boundary}=-\I^{(0)}_{\rm bulk}=1-\sum_{d|k}
\frac{1}{d^{2}}\ .
\label{result}
\end{equation}
As anticipated, the boundary term has precisely the effect of
subtracting out the higher winding number sectors.
Finally, using \eqref{ibulk2} we obtain
$\I_{{\EuScript L}^2}=\I_{\rm bulk}+\I_{\rm boundary}=+1$.
This value of the ${\EuScript L}^2$-index is consistent with the existence of
exactly one bound-state at threshold for $k$ D0-branes when $k=k'$. As the
result holds trivially for $k=1$,  we may now extend it to all values of $k$
by induction.

\section{The $\text{IIB}'$ Bound-State Problem}

In this section we will review the problem of finding BPS saturated
monopole-dyon bound-states required by Montonen-Olive duality
in ${\cal N}=4$ supersymmetric Yang-Mills
theory with gauge group $SU(2)$ in four dimensions.
Before discussing the relevant application of index theory,
we review the standard formulation of the problem due to Sen
\cite{Sen}.

In the moduli space approximation, the spectrum of BPS
states with magnetic charge $k$ can be determined by
considering the spectrum of supersymmetric quantum mechanics
on the classical moduli space of $k$
BPS monopoles \cite{gaunt,blum}, $\ms_{k}$. This space has the
isometric decomposition,
\begin{equation}
\ms_{k}= {\mathbb R}^{3} \times \frac{S^{1}\times \widehat\ms_{k}}
{{\mathbb Z}_{k}}\ ,
\label{decomp}
\end{equation}
where ${\mathbb R}^{3}$ is parametrized by the COM
coordinates
of the monopoles $\vec{X}$ and $S^{1}$, parametrized by
$\theta\in [0,2\pi]$, corresponds global charge
rotations of the configuration. The relative degrees of freedom
of the $k>1$ monopoles are coordinates on a smooth,
hyper-K\"{a}hler manifold, $\widehat\ms_k$, of real dimension
$d=4(k-1)$. The hyper-K\"{a}hler metric on $\widehat\ms_k$ is
only known explicitly in the $k=2$ case.
The ${\mathbb Z}_{k}$ reflects the invariance of a
multi-monopole configuration under a $2\pi$ charge angle of each
constituent monopole.

A bound-state in supersymmetric quantum mechanics is a state whose
wavefunction is non-normalizable as a function of its
COM coordinates. For example, a bosonic state of
definite COM momentum $\vec{P}$, has wavefunction
which goes like $e^{i\vec{P}\cdot\vec{X}}$ just like that
of a particle without substructure. On the other hand, the
part of the wavefunction which depends on the relative
coordinates must be normalizable. Furthermore, the quantum
mechanics of the COM degrees-of-freedom saturate the BPS bound
in the sector of charge $k$ for each value of $p$, which means
that the relative component of the bound-state wavefunction does
not contribute to the mass of the state. Hence the key
question for determining the bound-state spectrum is the
existence of normalizable zero energy states in supersymmetric
quantum mechanics on $\widehat\ms_k$.
If we introduce real coordinates $Y_{q}$ with
$q=1,2,\ldots , d=4(k-1)$ on $\widehat\ms_k$
the effective action takes the form,\footnote{All notation
is as in \cite{DKM97}.}
\begin{equation}
S^{(k)}= \tfrac12 \int dt \left[
g_{pq} d_{\tau} Y^p d_{\tau} Y^q +
g_{pq}  i \bar{\alpha}^p \gamma^0 D_{\tau} \alpha^q +
 \tfrac{1}{12}R_{pqrs} (\bar{\alpha}^{p}  \alpha^{r})
(\bar{\alpha}^{q}  \alpha^{s})\right]\ ,
\label{meff}
\end{equation}
where $\alpha_{q}$, are two-component fermionic
superpartners to the coordinates $Y_{q}$ with
$\bar{\alpha} = \alpha \gamma^0$ and $\gamma^0=\sigma^2$. Here,
$D_{\tau}\alpha^q = d_{\tau}\alpha^q +
d_{\tau} Y^r \Gamma^p_{rq}\alpha^q $ and $R_{pqrs}$
are the covariant derivative and Riemann curvature tensor formed
from the hyper K\"{a}hler metric metric $g_{pq}$.

The states we are looking for have a simple geometric
interpretation which was first described by Witten
\cite{Windex}. As usual the wavefunctions of purely
bosonic states just correspond to ordinary functions or zero
forms on $\widehat\ms_k$. The corresponding wavefunctions for
states of definite fermion number $F=r$ can be represented as
differential forms of degree $r$ on this manifold. Hence the
Hilbert space of the quantum mechanics \eqref{meff} is
essentialy the de Rham complex $\Lambda^{*}=
\oplus_{r=0}^{d} \Lambda^{r}$ of $\widehat\ms_k$. It will also
be useful to divide the Hilbert space into sectors of definite
charge $p$ under under the ${\mathbb Z}_{k}$ appearing in the denominator
of \eqref{decomp}, thus {\it e.g.\/}~$\Lambda^{(k)}_{r,p}$ is the space
of $C^{\infty}$ $r$-forms, $\omega_{r}$ on $\widehat\ms_k$
which transform as $\omega\rightarrow e^{2\pi ip/k} \omega$
under the generator of ${\mathbb Z}_{k}$.
As ${\mathbb Z}_{k}$ acts non-trivially
on the $S^{1}$ factor of the full moduli space
${\ms_{k}}$, the ${\mathbb Z}_{k}$ charge of states in
quantum mechanics on the relative moduli space $\widehat\ms_k$
is correlated with the electric charge of the resulting
BPS state. In fact a form with with ${\mathbb Z}_{k}$ charge $p$
corresponds to a state with electric charge $q=-p$ mod $k$.

By standard arguments the Hamiltonian of the quantum
mechanical system described above is precisely the Laplacian
$\Delta$ acting on forms. And zero energy states, being
zero modes of the Laplacian correspond to harmonic forms.
The existence of the $(q,k)$ dyons required by Montonen-Olive duality
then requires the existence of a unique, normalizable, harmonic
form of ${\mathbb Z}_k$ charge $p$ on the relative moduli space
$\widehat\ms_k$ only when $\langle p,k \rangle =1$.\footnote{Here and in
the following
$\langle m, n \rangle$ denotes the highest common divisor of two
integers $m$ and $n$.} As the harmonic
form
must be unique it must be invariant under Hodge duality
which maps $r$ forms to $d-r$ forms, preserving the
${\mathbb Z}_k$ charge and normalizability. This means that the
resulting forms must be middle dimensional, ie they must
be forms of degree $r=d/2=2k-2$. They must also be
(anti-)self-dual. As the hyper-K\"{a}hler metric on
$\widehat\ms_{2}$ has been determined explicitly by Atiyah
and Hitchin, the existence of the
required normalizable, self-dual, harmonic two-form on
this manifold can be established by a direct calculation
\cite{Sen}. The metric is unknown for $k>2$ and the statement
that the corresponding harmonic forms exist is known as
Sen's conjecure.

As described in the introduction, the problem of finding
the $(q,k)$ dyons is directly related to that of finding
BPS saturated $(q,k)$ strings in the IIB theory. In particular
we can consider a configuration where $k$ D-strings are stretched
between two parallel D3-branes. As in all our examples, the
world-volume dynamics of the D-strings are
described by a $U(k)$ gauge theory. The gauge theory is
formulated on a finite spacelike interval
between the two D3-branes and the effective theory is therefore
quantum mechanical at low energy. The resulting theory
includes adjoint scalar
fields whose eigenvalues describe the positions of the $k$
D-strings along the D3-world-volume. Dirichlet boundary conditions
for the D-string prevent motion in the transverse directions and
leave only eight unbroken supercharges. The world-volume theory also
has multiplets in the fundamental representation of $U(k)$ which are
localised at the endpoints of the interval. Electric charges can be
realized by including non-zero electric flux on the two dimensional
world sheet.

The $D$-term equations of the world-volume theory are
ordinary differential equations on the interval known as
Nahm's equations \cite{Diacon}. The boundary conditions at the
endpoints of the interval consititute the approporiate Nahm data
for the construction of exact $k$-monopole solutions in $SU(2)$ gauge
theory. The classical vacuum moduli space of the theory is obtained
by solving the $D$-term differential equations and modding out by the
$U(k)$ gauge symmetry. This is a Higgs branch in the sense that
the fields which acquire VEVs are adjoint hypermultiplets in the
language appropriate for a theory with eight supercharges.
This proceedure yields the classical moduli
space of $k$ BPS monopoles, denoted as $\ms_{k}$ above,
as an infinite dimensional hyper-K\"{a}hler quotient. At low-energy
the effective theory flows to a quantum mechanical
non-linear supersymmetric $\sigma$-model with target
$\ms_{k}$.
In other words, at low velocities we recover the standard moduli-space
description of BPS monopoles in the ${\cal N}=4$ theory described
above.

We will now briefly describe the elements necessary to discuss the
${\EuScript L}^2$-index theory relevant to the Sen bound state problem.
As above we must consider the ${\EuScript L}^2$-index of the Hamiltonian of
the $SU(k)$ supersymmetric gauge theory which describes the relative
motion of the D-strings. The new
feature is that the gauge theory has a non-singular effective
low-energy description in terms of
SUSY quantum mechanics on $\widehat\ms_k$ described by the Lagrangian
\eqref{meff}. Standard reasoning
suggests that it is sufficient to determine the presence or otherwise
of the required ground-states in the low energy theory.
As before we expect that the ${\EuScript L}^2$-index will include bulk and
boundary contributions.
\begin{equation}
\I^{(p,k)}_{{\EuScript L}^2}=\I^{(p,k)}_{\rm bulk}+
\I^{(p,k)}_{\rm boundary}
\label{split}
\end{equation}

As in the D0-brane example the bulk piece can
be extracted from the  effective action of the four
dimensional theory
compactified on a spacelike circle. As the $k$-monopole solution is
Bogomolnyi saturated it yields an instanton with eight exact
fermion zero modes.
The corresponding eight fermion term in the effective action
of the ${\cal N}=4$ theory on ${\mathbb R}^{3}\times S^{1}$ was recently
evaluated in \cite{part1,part2}, building on previous work in the
three-dimensional theory \cite{PSS}. The elements which enter this
calculation are very similar to those entering in the determination
of the exact ${\cal R}^{4}$ in the IIB effective action described
in the previous section. The requirements of supersymmetry force the
eight fermion term to obey Laplace's equation on the classical
Coulomb branch. Again the uniqueness properties of harmonic functions can be
exploited. In this case, the only additional assumption required is
the existence of a $Spin(8)_{\cal R}$ invariant superconformal fixed
point in three-dimensions.
This is in turn a direct consequence of IIA/M duality \cite{BFSS2}.
The final result is, $\I^{p,k}_{\rm bulk}=+1$
for all values of $p$ and $k$. Note that as in the IIA example the
bulk contribution to the index overcounts by precisely the sectors
where multiple winding can occur. In otherwords, if
$\langle p, k \rangle=l>1$ then we have an additional
instanton contribution which comes from the worldsheet of a
$(p/l,k/l)$-string stretched between the two D3-branes and wrapped
$l$ times around the compact Euclidean time dimension. As in the
IIA example, we will use Yi's argument to show that
the boundary contribution precisely
subtracts out the contributions of the multiple winding sectors.

We will consider the possible asymptotic states in the
sector of the theory with magnetic charge $k=k'$ and
electric charge $q=-p$ mod $k$. As before we will assume that the
Sen's conjecture holds for all $k<k'$. The simplest case is that
electric charge divisible by $k'$, $q=k's$.
In this case we have $k'$ dyons of electric charge $s$.
For the purposes of evaluating the boundary contribution
we will treat these as free massive non-relativistic particles moving
in three-dimensional space. As before the boundary contribution to the
index of the free system is minus the bulk contribution:
$\I^{(0)}_{\rm boundary}=-\I^{(0)}_{\rm bulk}$.
The later contribution can be evaluated using \eqref{result}.
As we have particles moving in three-dimensional space now with eight
superpartners the relevant values of $b$ and $f$ are 3 and 8
respectively, which yields $\I^{p,k}_{\rm boundary}=-\I^{(0)}_{\rm bulk}=-1$.

Now suppose we have non-zero electric charge $q=-p$ mod $k$
and that $k$ and $q$ have lowest common divisor
$l$. If $l$ is one then there are no asymptotic states in this sector
and $\I^{p,k}_{\rm boundary}=0$
If $l>1$, we get an asymptotic state of $l$
identical dyons of charges $(k/l,q/l)$. Again we will
treat these as $l$ identical particles moving in three dimensional
space with eight fermionic superpartners. This yields,
$\I^{p,k}_{\rm boundary}=-1$. Hence finally we obtain,
\EQ{
\I^{(p,k)}_{\rm {\EuScript L}^2}=\I^{(p,k)}_{\rm bulk}+
\I^{(p,k)}_{\rm boundary}  =\begin{cases} +1\qquad &
\langle p,k \rangle=1 \\0
&\langle p,k \rangle >1\ ,\end{cases}
\label{split2}
}
in accordance with Sen's conjecture.

There is also an interesting relation between the above results and
the topological properties of the moduli space of BPS monpoles.
To see this we consider the ${\EuScript L}^2$-index of supersymmetric
quantum mechanics on the moduli space $\widehat\ms_k$ without
implimenting the ${\mathbb Z}_k$ quotient. The relevant ${\EuScript L}^2$
Witten index counts
all normalizable harmonic forms on $\widehat\ms_k$ regardless of
${\mathbb Z}_k$ charge and is equal to
$\sum_{p=0}^{k-1} \I^{(p,k)}_{{\EuScript L}^2}$. We now consider the
path integral formula for this index. This is just a Euclidean path
integral with action obtained from (\ref{meff}) by Wick rotation
evaluated with periodic boundary conditions on the coordinates and
their superpartners. Taking the $\beta\rightarrow 0$ limit we obtain
the formula \cite{AG},
\begin{equation}
\sum_{p=0}^{k-1} \I^{(p,k)}_{\rm bulk}=
\frac{1}{(8\pi)^{d/2}(d/2)!} \int_{\widehat\ms_k}
\varepsilon^{p_{1}p_{2}\ldots
p_{d}}\ \varepsilon^{q_{1}q_{2}\ldots
q_{d}}\ R_{p_{1}p_{2}q_{1}q_{2}}\ldots R_{p_{d-1}p_{d}q_{d-1}q_{d}}\ .
\label{gbh}
\end{equation}
Thus the bulk contribution to the ${\EuScript L}^2$-index
can be represented as an integral of a $d=4(k-1)$ form density over the
moduli space $\widehat\ms_k$. Our previous results imply that this
quantity is equal to $k$. In the case $k=2$, where the explicit metric
on the moduli space was given by Atiyah and Hitchin one may
evaluate this integral explicitly and confirm that it equals two
\cite{GH}. To explain the significance of this result (for all $k$)
we review some elementary facts about classical index theory
which is distinct from the ${\EuScript L}^2$-index theory described above in
non-compact cases.

Classical index theory relates the zero mode spectrum of an elliptic
differential operator to the topological invariants of the
manifold on which it is defined. In the following we will be
interested in the index theorem for the Laplacian $\Delta$
acting on differential forms. The theorem takes it easiest to
state in the simple (but unrealistic) case of a
compact, smooth, manifold without boundary, $\ms$, of
real dimension $d$. In this case the
Laplacian will have a discrete spectrum and all its
eigenfunction are normalizable. The classical index of
$\Delta$ is identical to the ${\EuScript L}^2$-index in this simple case.

We denote the restriction of
the Laplacian on $\ms$ to the forms of degree $r$, as
$\Delta_{r}$. The number of linearly independent
harmonic forms of degree $r$ is therefore
$b_{r}={\rm dim}\, {\rm ker} \Delta_{r}$. The integers
$b_{r}$, for $0\leq r \leq d$ are known as the
Betti numbers of $\ms$. By Hodges theorem the Betti number
$b_{r}$ is also equal to the dimension of,
$H^{r}(\ms)$, the $r^{\rm th}$ cohomology group of $\ms$,
and are therefore purely topological (in otherwords they do
not depend of the choice of metric on $\ms$). The index,
$\chi$ of the Laplacian $\Delta$ is defined as the alternating
sum of the Betti numbers, $\chi= \sum_{r=0}^{d} (-1)^{r}b_{r}$.
Thus $\chi$ is just the Euler characteristic of $\ms$ and
in this well behaved case we have
\begin{equation}
{\rm Tr} (-1)^{F}\,e^{-\beta\H}={\rm Ind}\,\Delta=\chi\ .
\label{wb}
\end{equation}
The corresponding index theorem for the Laplacian is the
Gauss-Bonnet-Chern (GBC) theorem,
\begin{equation}
\chi={\rm Ind}\,\Delta = \int_\ms
e\left(T^{*}\ms\right)\ ,
\label{gbcompact}
\end{equation}
where the Euler density $e\left(T^{*}\ms\right)$
is defined by
\begin{equation}
e\left(T^{*}\ms\right)= \frac{1}{(8\pi)^{d/2}(d/2)!}
\varepsilon^{p_{1}p_{2}\ldots
p_{d}}\ \varepsilon^{q_{1}q_{2}\ldots
q_{d}}\ {R}_{p_{1}p_{2}q_{1}q_{2}}\ldots
{R}_{p_{d-1}p_{d}q_{d-1}q_{d}}\ .
\label{gbh2}
\end{equation}
This is precisely the integral appearing on the RHS of \eqref{gbh}.
In fact, in the compact case,
one may prove the Gauss-Bonnet theorem directly from
$\beta\rightarrow 0$ limit of the path integral \cite{AG}.

The relative moduli space of $k$-BPS monopoles is not compact
and so the GBC theorem stated above does not apply. To regulate
the problem one must replace the manifold $\widehat\ms_k$
by a compact manifold $\widehat\ms_k^{\rm cpt}$  with boundary
$\partial\widehat\ms_k^{\rm cpt}$ of finite volume $V$. For a review of
classical index theory on a manifold with boundary see \cite{EGH}.
This
could be done, for example, defining $\widehat\ms_k^{\rm cpt}$ as
the submanifold of $\widehat\ms_k$ where the distance between
any pair of monopoles is less than or equal to a fixed
length $R$. Although this sounds similar to the ${\EuScript L}^2$-index
introduced above there is an important difference: in this
standard approach we also include finite-volume boundary conditions which
render the spectrum of $\Delta$ discrete. As all states are normalizable in finite volume, the resulting index typically overcounts the number
of normalizable
groundstates by including states whose normalization diverges in the limit of
infinite volume. The modified version of
the Gauss-Bonnet theorem which holds under these conditions reads \cite{EGH}
\begin{equation}
\chi={\rm Ind}\,\Delta= \int_{\widehat\ms_k^{\rm cpt}}
e\big(T^{*}\widehat\ms_k^{\rm cpt}\big) +
\int_{\partial\widehat\ms_k^{\rm cpt}} Q\ .
\label{gbcompact2}
\end{equation}
The second term on the right-hand side is a surface term which involves the
integral of the second fundamental form $Q$ over the boundary
$\partial\widehat\ms_k^{\rm cpt}$. The geometical Euler characteristic $\chi$
is defined as the alternating sum over the Betti numbers just as in
the compact case.
This will be an integer which is independent of the volume $V$.
In contrast, the bulk and boundary contributions may depend
on $V$ and will not be
integral in general. However, the individual terms will each be
finite in the $V\rightarrow \infty$ limit and we denote these
limiting values $\tilde{\chi}$ and
$\delta{\chi}$ respectively and the theorem then states
$\chi=\tilde{\chi}+\delta\chi$. In the case
of two BPS monopoles one may use the explicit metric on the
moduli space to show that $\delta\chi=0$ and
$\chi=\tilde{\chi}=2$. Our result \eqref{gbh} shows that
$\tilde{\chi}=k$ for all values of $k$ \cite{DKM97}. Note that the bulk
contribution to the Euler characteristic is the same as the bulk
contribution to the corresponding ${\EuScript L}^2$-index while the boundary
contributions to these two quantities are different.

The homology of the moduli-space $\widehat\ms_k$ has
been determined by Segal and Selby, using the description of
this manifold as a space of rational maps. On a non-compact manifold
homology is dual to cohomology with compact support.
This cohomology with complex coefficients,
$H^{*}(\widehat\ms_k)$, is divided into different sectors
according to the action of
the discrete symmetry group, ${\mathbb Z}_k$.
Let $H^{*}(\widehat\ms_{k})_{p}$ denote the cohomology with compact
support restricted
to the sector of forms with ${\mathbb Z}_k$ charge $p$.
Then the result of Ref.~\cite{Segal}
is that $H^{r}(\widehat\ms_k)_{p}$
has complex dimension one whenever
$r=2k-2\langle p,k \rangle$ and dimension zero otherwise.
The Euler characteristic of $\widehat\ms_k$ can be deduced directly from
the results of Segal and Selby described in the previous section.
As the non-vanishing de Rham cohomology groups are of even dimension,
the Euler characteristic
is simply obtained by counting the total number of
solutions of the condition $r=2k-2\langle k,p \rangle$:
\begin{equation}
\chi(\widehat\ms_k)=\sum_{r=0}^{4(k-1)} \ \sum_{p=0}^{k-1}
\delta_{r,2k-\langle k,p \rangle} = k\ .
\label{resultss}
\end{equation}
This agrees with our expectations as long as $\delta \chi=0$.
As mentioned above, one may check this explicitly in the case
$k=2$.

\section{The $\text{IIA}'$ Bound-State Problem}

In this section we will consider the ${\EuScript L}^2$-index theory relevant
for determining the existence of bound-states of Yang-Mills
instantons thought of as solitons in five-dimensional gauge theory.

We begin by discussing the Type IIA
theory on flat ${\mathbb R}^{9,1}$ with $N$ D4-branes. The theory on the
D4-world-volume is a five dimensional $U(N)$ gauge theory with
sixteen supercharges. Dimensional reduction of this theory in one
direction yields ${\cal N}=4$ SUSY Yang-Mills theory in four
dimensions. As described above, the IIA theory also contains
D0-branes which are believed to form bound-states at threshold.
The D0-branes appear as Bogomol'nyi saturated solitons on the
five-dimensional world-volume of the D4-branes. These static
field configurations which have finite energy in five dimensions
correspond to instantons of finite Euclidean action in four
dimensions.

The theory on the D0-brane world volume is a quantum mechanical gauge
theory with eight supercharges. As in the absence of D4-branes,
this theory contains the same degrees of freedom as any
$U(k)$ gauge theory with sixteen supercharges (e.g. the ${\cal N}=4$ theory in
four dimensions). In particular, as in the basic IIA example,
the D0-world-volume theory contains nine scalar
fields in the adjoint representation of $U(k)$. The classical vacuum
manifold includes a Coulomb branch where the adjoint scalars have
non-zero VEVs breaking the gauge group down to its Cartan
subalgebra. The eigenvalues of these fields
describe the positions of the $k$ D0-branes
in their nine transverse directions.
However, the presence of D4-branes leads to
additional hypermultiplets on the D0-world-volume which transform
in the $(\Bk,\BN)$ representation of $U(k)\times U(N)$.
These multiplets break half the supersymmetry of the $N=0$
case and the corresponding scalars parametrize a new branch
of the vacuum moduli space (the Higgs branch) on which the gauge group
is broken completely. The resulting soliton configurations on the
D4-brane correspond to $k$ finite size instantons of gauge group
$U(N)$. The Higgs branch of the D0-brane theory
coincides with the moduli space of instantons in four dimensions.
Unbroken
supersymmetry and the Atiyah-Singer index theorem dictates
that this is a hyper-K\"{a}hler manifold of real dimension
$4kN$. As usual in a theory with eight supercharges,
we obtain the Higgs branch by imposing the $D$-term equations and
dividing out by $U(k)$ gauge transformations.
In the present case, this standard proceedure coincides with the
hyper-K\"{a}hler quotient construction of the corresponding instanton
moduli space $\ms_{k,N}$, which is also
known as the ADHM construction \cite{ADHM}.
In more detail, it is convenient to specify the
field content of the theory in the language of $d=4$
superfields. First of all, there is a vector multiplet of $\N=2$
supersymmetry,
decomposing as a vector multiplet $V$ of $\N=1$ SUSY and an adjoint chiral
multiplet $\Phi$. On top of this, there is an adjoint hypermultiplet,
consisting of chiral multiplets
$X$ and $\tilde X$, and $N$ fundamental hypermultiplets, consisting of $N$
chiral multiplets
$Q$ and $\tilde Q$.\footnote{We choose the $U(N)$ flavour symmetry,
which corresponds to the gauge symmetry on the D4-branes, to act as
$Q\to QU^\dagger$ and $\tilde Q\to U\tilde Q$.} The scalar components
of $\Phi$, which we denote with the same symbol, along with the 3
scalars that appear after dimensional reduction of the gauge field,
represent
the positions of the D0-branes transverse to the D4-branes. The
adjoint hypermulitplet $(X,\tilde X)$ specifies the positions of the
D0-branes within the D4-branes. On the Higgs branch, the D0-branes lie
within the D4-branes, {\it i.e.\/}~$\Phi=0$, and the
$D$-flatness conditions are
\AL{
&Q\tilde Q+[X,\tilde X]=0\ ,\label{adhm1}\\
&QQ^\dagger-\tilde Q^\dagger\tilde Q+[X,X^\dagger]+[\tilde X,\tilde
X^\dagger]=0\ ,
\label{adhm2}
}
respectively. These are the ADHM equations which, modulo the $U(k)$
gauge symmetry,
specify the moduli space $\ms_{k,N}$. The variables $Q$ and
$\tilde Q$ encode the instanton sizes and orientation in the $U(N)$
gauge group, while, as stated above, $(X,\tilde X)$ specify the
positions of the instantons in the D4-branes.
The $U(k)$ gauge theory also has Coulomb, as well as mixed, branches
which classically are connected to the Higgs branch at points where an
instanton shrinks to zero size, {\it i.e\/}~where some component of $Q$ and
$\tilde Q$ goes to zero and a $U(1)\subset U(k)$ does
not act freely. At these points, the D0-brane can move off the
D4-brane into the bulk and one moves out along another branch of the theory.

The terms in the Lagrangian of the theory can be grouped into three:
\EQ{
{\EuScript L}=g_1^{-2}{\EuScript L}_{V,\Phi}+{\EuScript L}_{X,\tilde X}+{\EuScript L}_{Q,\tilde
Q}\ ,
\label{lag}
}
where ${\EuScript L}_{V,\Phi}$ is the Lagrangian for the vector
multiplet, while ${\EuScript L}_{X,\tilde X}$ and ${\EuScript L}_{Q,\tilde
Q}$ are the Lagrangians describing the hypermultiplets. The vector
multiplet involves the dimensionful coupling constant
$g^{2}_1\sim (\alpha')^{-3/2}$.
Na\"\i vely, at energy scales much less
than $g_1^{2/3}$, the first term in \eqref{lag} is
irrelevant. Without a kinetic term, the vector multiplet becomes
non-dynamical and can be integrated out. The 3 auxiliary fields in $V$ impose
the ADHM constraints \eqref{adhm1} and \eqref{adhm2},
while the gauge field enforces $U(k)$
invariance. In the infra-red, therefore,
the gauge theory flows to
an effective theory which corresponds to the supersymmetric quantum
mechanics of a non-relativistic particle moving geodesically on
the ADHM moduli space $\ms_{k,N}$. This effective description
is a supersymmetric version of Manton's moduli space approximation
for soliton dynamics. As in the case of BPS monopoles the metric is
in curved space and hence there are non-trivial interactions between
solitons at order $v^{2}$ in the velocity expansion. However, an
important difference with this case is the fact that the metric has
orbifold singularities at the points where the classical Higgs phase
meets the mixed/Coulomb phases,
{\it i.e.\/}~where, as mentioned above, instantons shrink to zero size.
In this respect the problem
has common features to the basic problem of D0-binding in the IIA
theory. In particular, the moduli space approximation breaks down near
the singularities due to the presence of new light degrees of
freedom and therefore may not be
reliable for determining the existence of bound-states. The story of
the phases structure of these theories, both in quantum mechanics, as
here, and in $1+1$ dimensions, where one is concerned with the D1/D5
system, is a very interesting one (see
\cite{Witten1997,Berkooz1999,Aharony1999} and references
therein). For our
purposes the singularities in the Higgs branch
can be resolved in a standard way \cite{Nek} by
introducing non-commutivity in the five-dimensional gauge theory,
where the instantons appear as BPS solitons.
In terms of the string theory picture
this corresponds to introducing a background anti-symmetric tensor
field. In terms of the $U(k)$ gauge theory on the D0-brane world
volume, the non-commutivity parameters $(\zeta_{\mathbb
R},\zeta_{\mathbb C})$, corresponds to
Fayet-Illiopoulis terms in the center of the gauge group.
This coupling lifts the Coulomb branch of the world-volume theory and resolves
the orbifold singualrities of the Higgs branch, yielding a smooth
hyper-K\"{a}hler manifold $\ms_{k,N}^{(\zeta)}$ described by the
deformed ADHM equations:
\AL{
&Q\tilde Q+[X,\tilde X]=\zeta_{\mathbb C}1_{\sst[k]\times[k]}\ ,\\
&QQ^\dagger-\tilde Q^\dagger\tilde Q+[X,X^\dagger]+[\tilde X,\tilde
X^\dagger]=\zeta_{\mathbb R}1_{\sst[k]\times[k]}\ .
}

One of the most interesting features of the non-commutative
gauge theory on the D4-branes is that the theory admits non-singular
instantons even in the abelian case $N=1$. In fact,
the ADHM construction of the moduli space of $U(1)$ instantons as a
hyper-K\"{a}hler quotient is well defined even when the
non-commutivity parameter is zero. The resulting moduli space is
simply a symmetric product $\ms_{k,1}=\text{Sym}_{k}({\mathbb R}^4)$ reflecting
the fact that abelian instantons, or D0-branes sitting on a
single D4-brane, correspond to identical pointlike objects on ${\mathbb R}^4$.
In the simplest non-trivial case of two such objects we have
$\ms_{2,1}={\mathbb R}^4\times {\mathbb R}^4/{\mathbb Z}_2$,
where the ${\mathbb R}^4$ factor
describes the COM degrees-of-freedom and the orbifold
${\mathbb R}^4/{\mathbb Z}_2$
describes the relative positions of the two identical objects.
As the metric is flat,
the instantons do not interact at order $v^{2}$ and we certainly
cannot hope to find bound-states in a na\"\i ve moduli space
approximation. However, when the FI couplings are turned on,
the orbifold singularities are resolved and we have a
smooth manifold of the form ${\mathbb R}^4\times
{\widehat\ms}_{k,1}^{(\zeta)}$ where the manifold of relative positions
${\widehat\ms}_{k,1}^{(\zeta)}$ has non-zero curvature.
In the $k=2$ case the
orbifold ${\mathbb R}^4/{\mathbb Z}_2$ is
replaced by the Eguchi-Hanson manifold, as has been shown explicitly
in \cite{LeeTong}.

Once the singularities of the moduli space have been resolved,
the IR limit of the $U(k)$ gauge theory which describes the
relative dynamics of the solitons is simply a supersymmetric
$\sigma$-model with target space $\widehat\ms_{k,1}^{(\zeta)}$
of precisely the same form \eqref{meff} as occurs in the
BPS monopole case. In other words, the moduli space approximation
is now perfectly well defined and should yield sensible answers.
As in the monopole case, the required bound-states
precisely correspond to normalizable harmonic forms of middle
dimension on $\widehat\ms_{k,1}^{(\zeta)}$.
In the case $k=2$, the fact that the
Eguchi-Hanson manifold there is a single such form in agreement
with the prediction of the existence of single threshold bound state
of two abelian instantons \cite{LeeTong}. We will now discuss
the index theory responsible for establishing the existence of
required threshold bound states for all values of $k$. As before
the ${\EuScript L}^2$-index is the sum of bulk and boundary pieces
${\I}_{{\EuScript L}^2}={\I}_{\rm bulk}+{\I}_{\rm boundary}$
The main points are as follows:

{\bf 1:} The bulk contribution has two equivalent
representations. Firstly, by dimensionally reducing the partition
function of the world-volume gauge theory, it can be represented as the
partition function of a $U(k)$ matrix model with a single flavour
of matter in the fundamental representation. Secondly, by
dimensionaly reducing the low energy effective theory, it is equal
to the Gauss-Bonnet integral of the Euler density over the smooth
hyper-K\"{a}hler manifold $\widehat\ms_{k,1}^{(\zeta)}$.
The equality of these two
representations as ordinary integrals can be demonstrated by using
cohomological field theory techniques applied to the reductions of
supersymmetric gauge theories to matrix integrals, along the lines
described in \cite{MNS}. Actually,
the relevant point can already be seen in the
D1/D5 system \cite{partition}.
In that case, the kinetic term of the vector multiplet is
a $D$-term and hence $Q$-exact, where $Q$ is a nilpotent combination
of the supersymmetries \cite{Wittenphases}.
This fact descends to the matrix integral, and
allows one to argue {\it \'a la\/}~Moore {\it et al\/}~\cite{MNS},
that the resulting integral is independent of the gauge coupling
$g_0$. In particular, we can take the strong coupling limit
$g_0\to\infty$, {\em as long as the resulting integral is well
defined\/}. In the presence of non-commutativity, the FI terms ensures
that the resulting integral {\it is\/} well defined, since it is precisely the
integral of the Euler density over the smooth manifold
$\widehat\ms_{k,1}^{(\zeta)}$. In a similar way, one can argue that
with finite $g_0$, one can take the FI couplings to zero, since in
zero dimensions it is also $Q$-exact, without
altering the value of the integral. In other words, both
non-commutativity and $\alpha'$ effects,
through $g_0$, regulate the singular behaviour of the partition
function, a fact that was used in \cite{Aharony1998}.

{\bf 2:} As in the other examples we have considered the bulk
contribution to the ${\EuScript L}^2$-index can be related to an instanton
effect in a suitably compactified theory. In the present case,
the relevant effects are the D-instanton contributions to the
higher-derivative terms in the effective action of a single D3-brane of
the IIB theory.
Last year Green and Gutperle determined the order-${\alpha'}^4$ term
in the derivative expansion of the world-volume action.
In terms of the bosonic fields on the D3 world-volume,
it is given by \cite{GG2}
\EQ{S_{\sst D3}^{(4)\ \rm eff}={\alpha'}^4
{ \pi^3\over12}\int d^4x \sqrt{g^{(E)}}\,
h(\tau,\bar\tau)\Big((\partial^2
  \varphi)^4+ \tau_2(\partial^2\varphi)^2 \, \partial F^+\partial F^-
+\tau_2^2 (\partial F^+)^2
(\partial F^-)^2   \Big)
\ ,
\label{fourfmod}}
where $F_{mn}^{\pm}$ denote the (anti)-self-dual gauge field-strength,
$\phi_{AB}$ are the six real scalars of the ${\cal N}=4$ SYM
theory on the D3 world-volume, and $g^{(E)}$ is the Einstein metric.
The function $h(\tau,\bar\tau)$ was determined by considering the
$SL(2,{\mathbb Z})$ invariant
completion of a known tree level open string contribution
\EQ{S_{\sst D3}^{(4)\ \rm Born}=
{ \pi^3{\alpha'}^4\over12}\int d^4x \sqrt{g^{(E)}}\,
\tau_2 \Big((\partial^2
  \varphi)^4+ \tau_2(\partial^2\varphi)^2 \, \partial F^+\partial F^-
+\tau_2^2 (\partial F^+)^2
(\partial F^-)^2   \Big)
\ .
\label{highereinst}}
In order for the full nonperturbative
effective action to be modular invariant Green and Gutperle \cite{GG2}
proposed to replace the overall  factor of $\tau_2$ in \eqref{highereinst}
by a modular invarinat function, $h(\tau,\bar\tau)$,
\EQ{h(\tau,\bar\tau)= \ln|\tau_2 \,\eta(\tau)^4| \ ,
\label{hmodfn}}
where $\eta$ is the Dedekind function.  The
function $h$ has the weak coupling expansion,
\EQ{h(\tau,\bar\tau)= \Big(- {\pi \over 3}{\tau_2} +  \ln\tau_2- 2
  \sum_{k=1}^\infty
\sum_{d|k}{1\over d}\big( e^{2\pi i\ k \tau}+ e^{-2\pi i\ k \bar
\tau }\big)\Big)\ , \label{expanh}}
which  contains the Born contribution from \eqref{highereinst}, a
one-loop term
and an infinite set of multi-D-instanton and multi-D-anti-instanton
corrections.
In the Appendix
we will derive an expression for the generating functional $Z_{k,1}[\Phi]$
of the instanton-induced contributions to the scattering
amplitudes on the D3-brane. This generating functional will contain
an integral over the resolved centered instanton moduli space
as an overall coefficient.
This integral is precisely the bulk contribution to the
${\EuScript L}^2$-index on $\widehat\ms_{k,1}^{(\zeta)}$.
By comparing the 4-point amplitudes obtained from $Z_{k,1}[\Phi]$
to those derived from $S_{\sst D3}^{(4)\ \rm eff}$, we find that
the result of Green and Gutperle \eqref{expanh},
along with the $Q$-exactness argument
described in {\bf1} above, imply
\begin{equation}
\I_{\rm bulk}(k,1)=\sum_{d|k} \frac{1}{d}\ .
\label{ib2}
\end{equation}
As before, the terms in the sum can be interpreted in the
IIA theory on ${\mathbb R}^9\times S^{1}$ as coming from the worldline
of a threshold bound-state of $k/d$ D0-branes wrapped $d$ times on
$S^{1}$.

{\bf 3:} As previously we can apply the Yi argument to determine
the boundary contribution. The details are identical to those given
for the basic IIA bound-state problem in \S2. The only
difference is that the D0's move in four dimensions and have eight
single component fermions as superpartners: thus we have
$b=4$ and $f=8$. We can now use \eqref{proj2}, and the discussion
following it, to deduce
\begin{equation}
\I_{\rm boundary}(k,1)=1-\sum_{d|k} \frac{1}{d}\ .
\label{ib3}
\end{equation}
Hence, as expected, the boundary term subtracts of the
contributions of sectors of multiple winding to give results
consistent with the existence of a single threshold bound-state for
each value of $k$.

{\bf 4:} As in \S3, the bulk contribution to the
${\EuScript L}^2$-index is also equal to the bulk contribution
Euler characteristic $\chi_{k}$ of $\widehat\ms_{k,1}^{(\zeta)}$ in the
Gauss-Bonnet theorem. Thus we have
$\chi_{k}=\tilde{\chi }_{k}+\delta\chi_{k}$ with
$\tilde{\chi }_{k}=\sum_{d|k} d^{-1}$. Interestingly the full
Euler characteristic has recently been determined by Nakajima using
equivariant Morse theory \cite{Nak}. His result is
$\chi_{k}=\sigma(k)$ where $\sigma(k)$ is the
number of partitions of $k$. Thus we have a new
(and non-zero) prediction for the boundary contribution to the
Gauss-Bonnet theorem on $\widehat\ms_{k,1}^{(\zeta)}$:
\begin{equation}
\delta\chi_{k}=\sigma(k)-\sum_{d|k} \frac{1}{d}
\label{nak}
\end{equation}
This can be checked explicitly for the $k=2$ case.
As explained in \cite{EGH}, the Eguchi-Hanson manifold has
Euler characteristic $\chi_{2}=2$. This agrees with Nakajima \cite{Nak} as
$\sigma(2)=2$. The bulk contribution to the Gauss-Bonnet
theorem was evaluated in \cite{EGH}
using the explicit metric to obtain
$\tilde{\chi}_{2}=\tfrac32$ in agreement with points {\bf 2} and {\bf
4} above.
Finally the evaluation of the boundary term as an explicit
integral of the second fundamental form over the
$S^{3}/{\mathbb Z}_2$ boundary at infinity of the Eguchi-Hanson
manifold gives \cite{EGH} $\delta\chi_{2}=\tfrac12$
in agreement with \eqref{nak}.

It would be interesting to generalize this analysis
to the case of non-abelian instantons, $N>1$. Some results
for the corresponding theory with one compactified dimension were
presented in \cite{Lee},
indicating the existence of $N$ threshold bound-states for arbitrary $N$.
Applying our index theory approach to the non-compact theory
is complicated by the fact that that, for $N>1$, instantons have a
genuine size modulus. This means that the notion of a clustering
region is more complicated since even though instantons can be
spatially a long way apart they can still overlap by being large.
Nevertheless, although we have not calculated the bulk contribution to
the ${\EuScript L}^2$-index for general $N$ and $k$, we have calculated some
special cases which may offer some clues \cite{partition}.
Firstly, the bulk contribution to the ${\EuScript
L}^2$-index on $\widehat\ms_{1,N}^{(\zeta)}$ is
\EQ{
\I_{\rm bulk}(1,N)=\frac{2^{1-2N}(2N)!}{N!(N-1)!}\ .
\label{dff}
}
The case with $N=2$ is somewhat special, since the
unresolved (centered) moduli space is $\widehat\ms_{1,2}={\mathbb
R}^4/{\mathbb Z}_2$, which is the same as $\widehat\ms_{2,1}$. The
interpretation, however, is different: the radial parameter is the
scale size of the instanton and the $S^3$ solid angle gives the
$SU(2)$ orientation of the instanton (the ${\mathbb Z}_2$ orbifolding
corresponds to the center of the gauge group). As in the abelian case
of two instantons, the resolved space
$\widehat\ms_{1,2}^{(\zeta)}$ is also the Eguchi-Hanson manifold and
so $\I_{\rm bulk}=\tfrac32$, which agrees with \eqref{dff} for $N=2$. In
this case we also know that the boundary contribution to the ${\EuScript
L}^2$-index is $-\tfrac12$, since it is the same as for the $k=2$ and $N=1$
case. Hence, $\I_{{\EuScript L}^2}(1,2)=1$. For $N>2$, the
situation is not so simple because $\ms_{1,N}^{(\zeta)}$ does not have
a clustering region where it looks
like ${\rm Sym}_N({\mathbb R}^4)/{\mathbb R}^4$,
and we cannot, at least in any obvious way,
apply the Yi argument to these situations.

The situation for $k>1$ and $N>1$ is even more complicated. In this
case the only data that we have is the
bulk contribution to the ${\EuScript L}^2$-index of
$\widehat\ms_{k,N}^{(\zeta)}$ in the limit where $k$ is fixed and $N$
is large \cite{partition}:
\EQ{
\I_{\rm bulk}(k,N)
\underset{N\to\infty}=2^{3-2k}\pi^{6k-13/2}\sqrt
Nk^{3/2}\sum_{d|k}\frac1{d^2}\ .
}

\section{The IIB Bound-State Problem}

We now return to the IIB theory in ten flat dimensions. The problem is
to prove the existence of a unique bound-state of $k$ D-strings and $q$
fundamental strings whenever $k$ and $q$ are coprime and not
otherwise. The worldvolume theory is a two dimensional $U(k)$ gauge
theory with sixteen supercharges analysed by Witten in
\cite{bstates}. He started by wrapping the spatial dimension on a
circle so that we have quantum mechanics as in the other examples.
Fundamental string charge
corresponds to a constant electric flux, which is the same as a
charge at spatial infinity in this world sheet theory. The aim as before
is to prove the existence of appropriate
normalizable SUSY ground states of the corresponding $SU(k)$ gauge
theory which describes the relative degrees of freedom of the
D-strings. The
fundamental string charge then shows up as charge under the ${\mathbb Z}_k$
center of $SU(k)$ placed at infinity.
The coprime case is easy because the non-zero ${\mathbb Z}_k$ flux produces a
mass gap in the $SU(k)$ theory. The existence of the required
ground-state can then be confirmed reliably by a suitable
perturbation\cite{bstates}.

The non-coprime case is
much harder because it involves bound-states at threshold. However
the relevant ${\cal L}^2$-index theory can be studied using the
methods discussed in the preceeding sections.
In the following, we will only present a brief sketch of how we believe this
works. As in the other examples, the bulk
contribution to the index is obtained by considering the appropriate instanton
contributions for the IIB theory on ${\mathbb R}^{8}\times T^{2}$. The
instantons in question are just the $(q,k)$ strings wrapped on
$T^{2}$. These have sixteen zero modes and contribute to the
${\cal R}^{4}$ term in the IIB action of this compactified theory.
The exact ${\cal R}^{4}$ term was given by Kiritsis and Pioline,
extending the approach of Green and Sethi to the compactified theory.
The $(q,k)$ contribution is given in Eqn (3.46) of \cite{KP} as,
\begin{equation}
{\cal F}_{q,k}=-8\pi\, {\rm Re} \log \left[\prod_{n=1}^{\infty}
\left(1-\exp(2\pi
i n T_{q,k}) \right) \right]\ ,
\label{kp}
\end{equation}
where the action $T_{q,k}$ is the action of a single
wrapped $(q,k)$-string. If we take $T^{2}$ as a rectangular torus with
sides $R_{1}$ and $R_{2}$ and turn off the $B$ fields, then the action,
$T_{q,k}$, is the simply product of the world sheet area, $R_{1}R_{2}$, and
string tension $|q+k\tau|/\alpha'$.
The product over $n$ inside the logarithm yields a sum
over multiply-wrapped string world sheets for each type of string.

The relevant coefficients in the weak coupling expansion of (\ref{kp})
can be identified with
the R-R partition function of the $U(K)$ theory on a Euclidean
$T^{2}$. This has actually been calculated independently by Kostov and
Vanhove in \cite{Kostov}. In our notation, equation (33) of \cite{Kostov}
reads,
\begin{equation}
\Z_{K} \sim \sum_{d|K} \frac{1}{d} \sum_{Q=-\infty}^{+\infty}
\exp \left(-\frac{Q^{2}}{2 R_{1}R_{2}K}  \right)\ .
\label{kostov}
\end{equation}
The weak coupling limit of (\ref{kp}) agrees with (\ref{kostov})
if we identify $K=kd$ and $Q=qd$ with $d=n$.
To isolate the bulk contribution, we send $R_{1}$ and $R_{2}$ to zero and
obtain,
 \begin{equation}
{\cal I}_{\rm bulk}=\sum_{d|k} \frac{1}{d}\ .
\label{ib}
\end{equation}

As in the preceeding examples, the over
counting of the index corresponds
to the sectors with multiply wrapped branes and the
compensating boundary contribution can be evaluated using Yi's
argument as in the preceeding sections. In the present case
we can apply the argument by considering the winding modes of the
$(p,q)$ strings on ${\mathbb R}^{8,1}\times S^{1}$ as massive particles
in eight non-compact spatial dimension.
Now we may apply (\ref{proj2}) with the values
are $b=8$ and $f=16$, which yields $\delta=8-8=0$. Thus we have a
boundary contribution of $-1/d$ from each clustering sector with $d>1$
identical particles. The resulting index is unity when $k$ and $q$
are coprime and zero otherwise, as expected.

ND would like to thank Andrei Parnachev for useful discussions.

\startappendix

\Appendix{Bulk contribution to the ${\EuScript L}^2$-index on
$\widehat\ms_{k,1}^{(\zeta)}$}

In this Appendix we consider the generating functional
$\Z_{k,1}[\Phi]$
of the $k$-D-instanton contributions to the scattering
amplitudes on a single D3-brane. The expression for $Z_{k,1}[\Phi]$
will contain an integral over the centered resolved $k$-instanton
moduli space $\widehat\ms_{k,1}^{(\zeta)}$.
Matching the 4-point amplitudes obtained from $Z_{k,1}[\Phi]$
with the predictions of \eqref{fourfmod}, \eqref{expanh}
we will derive the expression \eqref{ib2}
for the bulk contribution to the ${\EuScript L}^2$-index on
$\widehat\ms_{k,1}^{(\zeta)}$.

In order to determine the generating functional
$\Z_{k,1}[\Phi]$,
we will first consider the partition
function of the $k$D(-1)/$N$D3 brane system\footnote{We are ultimately
interested in the case of a single D3-brane, but our formalism applies to all values of $N$. In due course we will set $N=1$ to simplyfy final expressions.}
on the $k$D(-1) world-volume
\cite{MO3},
\EQ{
\Z_{k,N} = \int d\mu_{k,N} \, e^{-S_{k,N}}\ .
\label{finfin}}
Here the instanton integration measure $d\mu_{k,N}$ and action $S_{k,N}$
are over the instanton collective coordinates, and the D-3 brane
degrees of freedom are turned off.
Next, we will turn on the fields living on the world-volume
of the D3-branes.
>From the point of view of the zero-dimsional
instanton world-volume, the D3-branes are infinitely heavy and their
world-volume fields are represented by static sources $\Phi$ coupled to
instanton collective coordinates in the action $S_{k,N}(\Phi)$.
The partition function \eqref{finfin}
in the presence of the D3-sources $\Phi$,
\EQ{
\Z_{k,N} [\Phi] = \int d \mu_{k,N} \, e^{-S_{k,N} (\Phi)}\ ,
\label{finfff}}
defines the generating functional
of the D-instanton-induced amplitudes on the D3-branes in the semi-classical
approximation.

The $k$D$p$/$N$D$(p+4)$ brane system can live in the maximal dimension
$p=5$ which corresponds to the 6-dimensional gauge theory on the world-volume
of the D5-branes. Then the cases $5\ge p \ge -1$ follow by dimensional reduction. In Section 4 we specified the ADHM collective coordinates in the
language of $d=4$ superfields. This corresponds to choosing
$p=3$ as the starting point of dimensional reduction to $p=-1$.
For practical calculations it is more convenient to start with the maximal
case $p=5$ and follow conventions of Ref.~\cite{MO3}.
The field content of the $k$D5/$N$D9 system is described by the
$(1,1)$ vector multiplet and two bi-fundamental hypermultiplets in the
6-dimensional world-volume of $k$D5 branes. The component fields are
introduced in the same way as in \cite{MO3} and are listed in the
Tables 1-2.
\begin{table}\setlength{\extrarowheight}{5pt}
\begin{center}\begin{tabular}{||l|l|l|r||} \hline\hline
\phantom{$\Biggr($}
Component & Description & $U(k)$ & $U(N)$ \\
\hline\hline
$\quad\chi^{1 \ldots 6}$ & Gauge Field & ${\Bk}\times{\Bk}$
& ${\bf 1}\quad$
\\
$\quad\lambda_{\aD}$ & Gaugino & ${\Bk}\times{\Bk}$
& ${\bf 1}\quad$
\\
$\quad D^{1 \ldots 3}$ & Auxiliary Field & ${\Bk}\times{\Bk}$
& ${\bf 1}\quad$
\\
\hline
$\quad a^{\prime}_{\alpha \aD}$ & Scalar Field & ${\Bk}\times{\Bk}$
& ${\bf 1}\quad$
\\
$\quad {\cal M}^{\prime}_{\aD}$ & Fermion Field & ${\Bk}\times{\Bk}$
& ${\bf 1}\quad$
\\
\hline\hline
\end{tabular}\end{center}
\caption{\small $(1,1)$ vector multiplet in $d=6$.}
\end{table}

\begin{table}\setlength{\extrarowheight}{5pt}
\begin{center}\begin{tabular}{||l|l|l|r||} \hline\hline
\phantom{$\Biggr($}
Component & Description & $U(k)$ & $U(N)$ \\
\hline\hline
$\quad w_{\aD}$ & Scalar Field & ${\Bk}$ & ${\BN}\quad$
\\
$\quad\mu$ & Fermion Field & ${\Bk}$ & ${\BN}\quad$
\\
\hline
$\quad\bar{w}_{\aD}$ & Scalar Field & ${\bar{\Bk}}$
& ${\bar{\BN}}\quad$
\\
$\quad\bar\mu$ & Fermion Field & ${\bar{\Bk}}$ & ${\bar{\BN}}\quad$
\\
\hline\hline
\end{tabular}\end{center}
\caption{\small Bi-fundamental hypermultiplets in $d=6$.}
\end{table}
The relation between the $d=4$ language of Section 4
and the $d=6$ language is as follows:
\EQ{
\chi^{1\ldots 4} \equiv V\ , \quad
\chi^5 \pm i \chi^6 \equiv \Phi^{(\dagger)} \ , \quad
a'_{\alpha\aD}\equiv\begin{pmatrix} X^\dagger & \tilde X\\
-\tilde X^\dagger & X \end{pmatrix} \ , \quad
w_\aD\equiv\begin{pmatrix}Q^\dagger \\ \tilde Q\end{pmatrix}\
 ,\quad \bar w^\aD\equiv\begin{pmatrix} Q & \tilde Q^\dagger
\end{pmatrix}\ .
\label{rell}}
The ${\cal N}=1$ superfelds on the right hand sides of
equations above are identified with their
bosonic components.

The D-instanton integration measure in the ${\cal N}=4$ supersymmetric $U(N)$ gauge theory has the following form \cite{MO3}:
\EQ{
\Z_{k,N} = \frac{g_4^4}{{\rm Vol}\,U(k)}
 \int d^{4k^2}a'\,
d^{8k^2}{\cal M}'\, d^{6k^2}\chi\, d^{8k^2}\lambda\, d^{3k^2}D\,
d^{2kN}w\,d^{2kN}\bar w\, d^{4kN}\mu\, d^{4kN}\bar{\mu}\
\exp [-S_{k,N}]
\label{final1}}
where $S_{k,N}=g_0^{-2}S_{G} + S_{K}+S_{D}$ and
\begin{subequations}
\begin{align}
S_{G} & = {\rm tr}_{k}\big(-[\chi_a,\chi_b]^2+\sqrt{2}i\pi
\lambda_{\dot{\alpha}A}[\chi_{AB}^\dagger,\lambda_B^{\dot{\alpha}}]
+2D^{c}D^{c}\big)\, ,\label{p=-1actiona} \\
S_{K} & =  -{\rm tr}_{k}\big([\chi_a,a'_{n}]^2
+\chi_a\bar{w}^\aD_{u}
w_{u\dot{\alpha}}\chi_a + \sqrt{2}i\pi
{\cal M}^{\prime \alpha A}[\chi_{AB},
{\cal M}^{\prime B}_{\alpha}]+2\sqrt{2} i \pi
\bar{\mu}_{u}^{A}\chi_{AB}\mu^{B}_{u}\big)\ ,\label{p=-1actionb} \\
S_{D} & =  i  \pi{\rm tr}_k\big(
[a'_{\alpha\dot{\alpha}},{\cal M}^{\prime\alpha A}]\lambda^{\dot{\alpha}}_{A}
+\bar{\mu}^{A}_{u}w_{u\dot{\alpha}}\lambda^{\dot{\alpha}}_{A}+
\bar{w}_{u\dot{\alpha}}\mu^{A}_{u}
\lambda^{\dot{\alpha}}_{A} + \pi^{-1}D^{c}(\tau^c)^\bD_{\ \aD}
(\bar w^\aD w_\bD+\bar a^{\prime\aD\alpha}a'_{\alpha\bD})\big)\ .
\label{p=-1actionc}
\end{align}\end{subequations}
We use the same conventions\footnote{The index assignement is
$i,j,\ldots=1,\ldots,k$ and $u,v,\ldots=1,\ldots,N$, together with
$A,B,\ldots=1,\ldots,4$ and
$a,b\ldots=1,\ldots,6$ and $c\ldots=1,\ldots,3.$
An overall normalization of the measure is suppressed in \eqref{final1}.
It
can be determined by comparison with the ADHM
multi-instantn measure in field theory. Properly normalized
centered partition function $\widehat\Z_{k,N}$
is discussed in the Appendix of \cite{partition}.}
as in \cite{MO3}. For future convenience we also introduce
the centre of mass coordinates of the $k$-instanton,
\EQ{
{a^{\sst CM}}_n=k^{-1}{\rm tr}_k\,a'_n\ , \quad
{{\cal M}^{\rm ss}}^A_{\alpha}= k^{-1}{\rm tr}_k
({\cal M}^{\prime A}_{\alpha}) \ .
\label{sscms}
}
These simply
correspond to translations along the D3-branes in superspae of the multi-instanton configuration as a whole.

The D-instanton partition function $\Z_{k,N}$
depends explicitly on the inverse string tension $\alpha'$
through the
zero-dimensional coupling $g^{2}_{0}\propto (\alpha')^{-2}$ which appears
in $g_0^{-2}S_{G}$ which comes from the dimensional reduction of the $d=6$
gauge action.
In the field theory limit the fundamental string scale is set
to zero, $\alpha'=0$, to decouple the world-volume
gauge theory from gravity. Thus, as explained in \cite{MO3},
to derive the ADHM-instanton measure
in conventional supersymmetric gauge theory
one must take the limit $\alpha'\rightarrow 0$.
In this limit $g^{2}_{0}\rightarrow \infty$ equations of motion
for $D^c$ are precisely the non-linear ADHM constraints
\eqref{adhm1}, \eqref{adhm2}.
Similarly equations of motion for $\lambda$ are the fermionic ADHM
constraints. Integration over $D^c$ and $\lambda^\aD_A$ yields
$\delta$-functions which impose the constraints.

For D-instanton applications
in string theory we must keep the $\alpha'$ corrections
in the D-instanton measure \eqref{final1}.
In this case, the
integrals over $D^c$ and $\lambda^\aD_A$ do not lead to the ADHM
constraints and their fermionic analogues; instead of being imposed
via delta functions, the arguments of the ADHM constraints appear as
Gaussian factors:
\EQ{
\int d^3D\,e^{-{\rm tr}_k(2g_0^{-2}D^cD^c+iD^c\B^c)}
=(g_0^2\pi/2)^{3k^2/2}e^{-\tfrac18
g_0^2{\rm tr}_k\B^2}
\ . \label{strcof}
}
This provides a natural string theory regularization of the
singularities of the instanton moduli space.
We also note that the $g_0^{-2}\propto {\alpha'}^2$
terms
in the action of \eqref{final1} lift the superconformal fermion zero modes.
When $\alpha'$ is set to zero in \eqref{final1} and the $U(N)$ gauge group
is {\it unbroken}, the superconformal fermion zero modes are exact.
(This is only relevant for $N>1$.)
In the latter situation to obtain non-zero answers one has to saturate
the supersymmetric as well as the superconformal fermion zero modes
by the field insertions in the pre-exponent as in \cite{MO3}, or otherwise.

We now need to include
on the right hand side of \eqref{final1} the interactions
with the D3-brane world-volume fields
transforming in the adjoint
represetation of $U(N)$: the $d=4$ gauge fields
$A_m(x)$, the six scalars $\varphi_{AB}(x)$, and the gauginos
$\Lambda^A_\alpha(x),{\bar\Lambda}_A^{\dot\alpha}(x)$.
Note that so far we have
ignored these fields, the bifundamental hypermultiplets $w$ and $\bar{w}$
represent merely the strings streched  between the D-instantons and the
D3-branes.
The gauge interactions between the D9-gauge fields,
$A_\mu=({\varphi_{AB},A_m}),$
and the D5-matter fields, $a'_m$, ${w}_{\dot{\alpha}}$ and $\bar{w}_{\dot{\alpha}},$
occur in covariant derivatives.
These interactions are introduced by an extension of the gauge sector in $d=6$
matter actions $S_K$ and $S_D$
\SP{
(\chi_{AB})_{ij} \ \delta^{uv} \ &\rightarrow\ (\chi_{AB})_{ij}\  \delta^{uv}
- \varphi^{uv}_{AB}(a^{\sst CM})\ \delta_{ij} \ , \\
(D^c)_{ij}\ \delta^{uv} \ &\rightarrow\ (D^c)_{ij}\ \delta^{uv}
- F^{+uv}_{mn}(a^{\sst CM}) \ \bar{\eta}^c_{mn}\delta_{ij} \ ,
\label{chg1}}
and keeping the gauge action $S_G$ unchanged.
Note that $S_K$ and $S_D$ are defined on the D-instanton world-volume
and do not depend on the D3 world-volume coordinates $x_m$. Thus,
the D3 world-volume fields in \eqref{chg1} are taken at the specific
point $x_m \equiv a^{\sst CM}_m$, the centre of mass location of the
instanton \eqref{sscms} along the D3 brane. This is consistent with the
earlier observation that from the instanton world-volume
point of view, the D3 fields
are infinitely heavy and should appear as static sources.\footnote{
We further note that for the general $N>1$ case
the separations
between the D3-branes are automatically incorporated in \eqref{chg1}
by turning on the VEVs of the scalar fields,
$\langle \varphi^a \rangle_{uv} = {\rm diag}_{uv} \{
\langle \varphi^a \rangle_1, \cdots ,\langle \varphi^a \rangle_N \}.$
When the separations are non-vanishig for all $u\neq v$ the $U(N)$
gauge group is spontaneously broken to $U(1)^N$ and we are on the
Coulomb branch of the $N$D3 system.}

However, the interaction term generated by the substitution
\eqref{chg1} is not the full story. With \eqref{chg1} we have introduced
an explicit dependence on the centre of mass instanton
coordinates $a^{\sst CM}_m$, but not on their fermionic superpartners
${\cal M}^{\rm ss}$. With respect to the (super)symmetry transformations
broken by D$(-1)$ branes, but not by D3-branes
the total action of the D$(-1)$/D3 system
$S_{k,N}^{\rm tot}$
should transform as follows:
\AL{a^{\sst CM}_m \rightarrow a^{\sst CM}_m + z_m \ : \quad
S_{k,N}^{\rm tot}\ (a^{\sst CM}_m,\  {\cal M}^{\rm ss}, \cdots)\
&\rightarrow \ S_{k,N}^{\rm tot}\ (a^{\sst CM}_m+ z_m,\
{\cal M}^{\rm ss}, \cdots) \ ,
\label{trtr1} \\
{{\cal M}^{\rm ss}}^{ A}_\alpha \rightarrow
{{\cal M}^{\rm ss}}^{ A}_\alpha + \eta^A_\alpha \ : \quad
S_{k,N}^{\rm tot}\ (a^{\sst CM}_m,\ {\cal M}^{\rm ss}, \cdots)\
&\rightarrow \ S_{k,N}^{\rm tot}\ (a^{\sst CM}_m,\
{\cal M}^{\rm ss}+\eta, \cdots) \ .
\label{trss1}
}
The factor $\int d^4 a^{\sst CM} \ d^8 {\cal M}^{\rm ss}$
in the generating functional
\eqref{finfff} will ensure that when the D$(-1)$ branes are integrated out,
the symmetries broken by instantons on the D3 branes will be restored.

After the substitution \eqref{chg1}, the action satsfies
\eqref{trtr1}, but not \eqref{trss1}.
An elegant way to satisfy \eqref{trss1} is to upgrade the bosonic fields
on the right hand sides of \eqref{chg1} to the corresponding components of the
$\N=4$ on-shell superfields ${\cal W}_{AB}(\theta,\bar\theta)$ \cite{Howe}.
Using supersymmetric covariant derivatives one can also define
${\cal W}_B^\aD  = (3/2) D^{A \aD} {\cal W}_{AB}$ and
${\cal W}_{mn} =  D^A \bar \sigma_{mn}  D^B {\cal W}_{AB}$.
The dependence on ${\cal M}^{\rm ss}$ is introduced as follows \cite{GG2}:
\EQ{\Phi_{mn} = {\cal W}_{mn}|_{\bar \theta =0, \theta = {\cal M}^{\rm ss}}\ ,
\qquad
\Phi_A^\aD = {\cal W}_A^\aD |_{\bar \theta =0, \theta ={\cal M}^{\rm ss}}\ ,
\qquad
\Phi_{AB} = {\cal W}_{AB}|_{\bar \theta =0, \theta = {\cal M}^{\rm ss}} \ .
\label{srcssf}}
By construction,
these superfields are  invariant under the supersymmetry transformations
generated by shifts of ${\cal M}^{\rm ss}\to {\cal M}^{\rm ss} +\eta$,
\EQ{\delta_{\eta} \Phi = \eta_\alpha^A {\partial \over \partial
 {{\cal M}^{\rm ss}}^{A}_\alpha } \Phi \ .
\label{susytrs}}
Thus, the total interaction with the $N$D3-brane sources is determined by
the substitution
\SP{
(\chi_{AB})_{ij} \ \delta^{uv} \ &\rightarrow\ (\chi_{AB})_{ij}\  \delta^{uv}
- \Phi^{uv}_{AB}(a^{\sst CM},{\cal M}^{\rm ss})\ \delta_{ij} \ , \\
(\lambda_A^\aD)_{ij} \ \delta^{uv} \ &\rightarrow\ (\lambda_A^\aD)_{ij}
 \ \delta^{uv} -\Phi^{uv\aD}_A (a^{\sst CM},{\cal M}^{\rm ss})\ \delta_{ij}
 \ , \\
(D^c)_{ij}\ \delta^{uv} \ &\rightarrow\ (D^c)_{ij}\ \delta^{uv}
- \Phi^{uv}_{mn}(a^{\sst CM},{\cal M}^{\rm ss}) \ \bar{\eta}^c_{mn}\delta_{ij}
 \ ,
\label{chg2}}
into the instanton matter-field actions $S_K$ and $S_D$.
This inclusion of the D3-brane sources from now on will be denoted as
$S_K(\Phi)$ and $S_D(\Phi)$.
Equation \eqref{chg2} was originally derived by Green and Gutperle \cite{GG2}
from slightly different considerations.
We will need an explicit realization of only one of these superfields
which we copy from \cite{GG2}
\SP{
\Phi_{mn}(a^{\sst CM}_n, {\cal M}) =& F^+_{mn} + i{\cal M}^{ A} \sigma_{[m} \, \partial_{n]}
\bar\Lambda_A +4 {\cal M}^B \sigma_{[m}^{\ \ p} {\cal M}^A \partial_{n]} \partial_p
\varphi_{AB}\\
& + 2 \epsilon_{ABCD} {\cal M}^B \sigma_{p[m}{\cal M}^A {\cal M}^C
\partial_{n]} \partial_p \Lambda \\
& + \epsilon_{ABCD}
{\cal M}^B\sigma_{p[m}{\cal M}^A\; {\cal M}^C \sigma^{kl}{\cal M}^D\; \partial_{n]}
\partial_p F^-_{kl}\ .
\label{sfd1}
}
The fact that an {\it on-shell} superfield is used in the above analysis is
not an obstacle for deriving instanton-induced amplitudes (which are
also on-shell quantities) on the 3-branes.

The $k$D(-1)/$N$D3 partition function on the $k$D(-1) world-volume and
in the presence of the D3-brane sources is given by
\SP{
\Z_{k,N}[\Phi] = \frac{g_4^4}{{\rm Vol}\,U(k)}\,
\int\,& d^{6k^2}\chi\, d^{8k^2}\lambda\, d^{3k^2}D\,d^{4k^2}a'\,
d^{8k^2}{\cal M}'\,
d^{2kN}w\,d^{2kN}\bar w\, d^{4kN}\mu\, d^{4kN}\bar{\mu}\ \\
&\times\exp -S_{k,N}(\Phi)\ ,
\label{finalphi}
}
where
\EQ{
S_{k,N}(\Phi)=g_0^{-2}S_{G} + S_{K}(\Phi)+S_{D}(\Phi)
\ . \label{sknphi}}
It is worthwhile to note that the bosonic zero modes associated with
4-translations $a^{\sst CM}$ and
the supesymmetric fermionic zero modes
${\cal M}^{\rm ss}$ are both lifted in the action \eqref{sknphi}
via explicit insertions of
the superfields $\Phi(a^{\sst CM},{\cal M}^{\rm ss})$.
There remain no unlifted bosonic or fermionic zero modes in
the partition function $\Z_{k,N}[\Phi]$.

The generating functional \eqref{finalphi} can be simplified
for the case of a single D3 brane. Thus from now on we set $N=1$.
The simplification comes about by noticing that the
superfields in the shifts
\eqref{chg2} of $S_K(\langle \varphi\rangle)$ and $S_D$
are now constants (not $[N]\times[N]$ matrices).
These constant shifts can be undone by the opposite constant
shifts in the integration
variables $\chi$, $\lambda$ and $D$ in the partition function
and in the previously unshifted action $S_G$. Furthermore,
since $\chi$ and $\lambda$ variables appear in $S_G$ only in the
commutators, the shifts of $\chi$ and $\lambda$ cancel and only
the shift of $D$ contributes:
\EQ{g_0^{-2}S_G\rightarrow g_0^{-2}
{\rm tr}_{k}
\big(-[\chi_a,\chi_b]^2+\sqrt{2}i\pi
\lambda_{\dot{\alpha}A}[\chi_{AB}^\dagger,\lambda_B^{\dot{\alpha}}]
+2(D^{c}+\Phi_{mn}\bar{\eta}^c_{mn})^2\big)\, .
\label{sgneww}
}
Consider now the terms in the total action involving the $D$-fileds:
\SP{&2g_0^{-2}(D^{c}+\Phi_{mn}\bar{\eta}^c_{mn})^2\big)+
iD^{c}(\tau^c)^\bD_{\ \aD}
(\bar w^\aD w_\bD+\bar a^{\prime\aD\alpha}a'_{\alpha\bD})\big) \ =\\
 &2g_0^{-2}(\Phi_{mn}\bar{\eta}^c_{mn})^2+2g_0^{-2}(D^{c})^2+
iD^{c}(\tau^c)^\bD_{\ \aD}
(\bar w^\aD w_\bD-i4g_0^{-2}\Phi_{mn}\bar{\eta}^c_{mn}
+\bar a^{\prime\aD\alpha}a'_{\alpha\bD})\big) \ .
\label{rearr1}}
Using this rearrangement we can finally express the generating
functional in a simple form:
\EQ{\Z_{k,1}[\Phi] \ =\
g_4^4\;\int\, d^4 a^{\sst CM} d^8 {\cal M}^{\rm ss} \
\exp[-2g_0^{-2}(\Phi_{mn}(a^{\sst CM},{\cal M}^{\rm ss})\,
\bar{\eta}^c_{mn})^2] \ \times \I_{k,1}
\ ,
\label{finalrr}
}
where $\I_{k,1}$ denotes an integral over the centered $k$-instanton
moduli space,
\SP{
\I_{k,1}\ =\  \frac{1}{{\rm Vol}\,U(k)}\,
&\int\,  d^{6k^2}\chi\, d^{8k^2}\lambda\, d^{3k^2}D\,d^{4(k^2-1)}a'\,
d^{8(k^2-1)}{\cal M}'\,
d^{2k}w\,d^{2k}\bar w\, d^{4k}\mu\, d^{4k}\bar{\mu}\\
&\exp[-g_0^{-2}S_{G} - S_K - S_D(\zeta)]
\ . \label{finaljj}
}
Here $S_G$ and $S_K$ are given by, correspondingly
\eqref{p=-1actiona} and \eqref{p=-1actionb}, and contain no D3-sources.
$S_D(\zeta)$ is given by \eqref{p=-1actionc} with the $D$-term
shifted by
$\zeta \equiv i4g_0^{-2}\Phi_{mn}\bar{\eta}^c_{mn}$,
\EQ{S_{D}(\zeta) =  i  \pi{\rm tr}_k\big(
[a'_{\alpha\dot{\alpha}},{\cal M}^{\prime\alpha A}]\lambda^{\dot{\alpha}}_{A}
+\bar{\mu}^{A}w_{\dot{\alpha}}\lambda^{\dot{\alpha}}_{A}+
\bar{w}_{\dot{\alpha}}\mu^{A}
\lambda^{\dot{\alpha}}_{A} + \pi^{-1}D^{c}(\tau^c)^\bD_{\ \aD}
(\bar w^\aD w_\bD-\zeta+\bar a^{\prime\aD\alpha}a'_{\alpha\bD})\big)\ .
\label{sdzeta}
}
Generally speaking, $\I_{k,1}$ is a certain function of two parameters: $\zeta$ and $g_0$, expressed as an integral \eqref{finaljj}
over the centered $k$-instanton
moduli space $\widehat\ms_{k,1}^{(\zeta)}$.
This integral is the partition function of the $U(k)$
instanton matrix model with a single flavour of matter in the fundamental
representation; $g_0$ is the $U(k)$ gauge coupling, and $\zeta$
is the abelian FI parameter. Both $g_0$ and $\zeta$ provide a resolution
of the singularities on $\widehat\ms_{k,1}$.
For $g_0<\infty$ and arbitrary $\zeta$ the singularities
of $\widehat\ms_{k,1}$ are absent due to a string-theory resolution
and the integral is well-defined. Alternatively, for $\zeta\neq 0$
and arbitrary $g_0$ the singularities
of $\widehat\ms_{k,1}$ are again absent due to a FI resolution\footnote{
As in Section 4 the $\zeta$-resolved moduli space
describes instantons in  gauge theory on a spacetime
with non-commuting coordinates \cite{Nek}.
Note, however,
that we started with the ordinary commutative theory on the 3-brane and
derived the
$\zeta$-resolution from the string-theory resolution via the inclusion
of the D3-brane sources:
when $\alpha'$ is set to zero,
$\zeta\propto g_0^{-2}\Phi \propto {\alpha'}^2\Phi$ is zero as well.},
and the integral $\I_{k,1}$ is well-defined again.
The Q-exactness argument of Section 4 implies that
$\I_{k,1}$ does not depend on
$g_0$ and does not depend on $\zeta$ and is precisely
equal to the bulk
contribution to the ${\EuScript L}^2$-index on the resolved centered
instanton moduli space $\widehat\ms_{k,1}^{(\zeta)}$.

The expression \eqref{finalrr} is the generating functional
for the instanton-induced scattering amplitudes. In particular,
the 4-point amplitudes ${\cal A}_{2 \to 2}$
generated by \eqref{finalrr} correspond precisely
to those derived from the effective action \eqref{fourfmod}!
A simple way to see this is to single out specific components of the
on-shell superfield $\Phi_{mn}(a,{\cal M})$. For example,
it is easy to show that the
$(\partial^2 \varphi)^2$ term in the effective action \eqref{fourfmod}
folows from the
$4 {\cal M}^B \sigma_{[m}^{\ \ p} {\cal M}^A \partial_{n]} \partial_p
\varphi_{AB}$ component in the expansion of $\Phi_{mn}(a,{\cal M})$ in
Eq.~\eqref{sfd1}.
The 4-point scalar amplitude is obtained from the generating functional
\eqref{finalrr} via the functional differentiation with respect to
the Fourier components of the scalar fileds $\tilde{\varphi}^a(k_i)$,
and multiplications by the polarizations $\zeta_i^a$,
\EQ{{\cal A}_{2 \to 2}^{\rm scalar}(k_1,k_2,k_3,k_4)
= {1\over 4!} \prod_{i=1}^4 \left(
\zeta_i^{a_i}{\delta \over \delta \tilde{\varphi}^{a_i}(k_i)}\right)\,
\Z_{k,1}[\Phi] \ ,
\label{4ptamp}}
and with the substitution of the selected superfield component
\EQ{\Phi_{mn}(a,{\cal M}) = 4 {\cal M}^B \sigma_{[m}^{\ \ p}
{\cal M}^A \partial_{n]} \partial_p \varphi_{AB}(a) \ .
\label{sbstn}}
For a non-zero answer we need to saturate precisely 8 supersymmetric
fermion zero modes ${\cal M}$, and the argument in the exponent of
\eqref{finalrr} needs to be brought down 8 times.
Schematically we have
\SP{
&\prod_{i=1}^4 \left(\zeta_i{\delta \over \delta \tilde{\varphi}(k_i)}\right)\,
\int d^4 a \int d^8{\cal M} \,{1\over 4!} \,
({\cal M}{\cal M}\, \partial\, \partial\, \varphi(a))^4  =\\
&\prod_{i=1}^4 \left(\zeta_i{\delta \over \delta \tilde{\varphi}(k_i)}\right)\,
\int d^4 a \int d^8{\cal M} \,{1\over 4!} \,
\left({\cal M}{\cal M} \int d^4p e^{-ip_m a^m} p\,p\,
\tilde{\varphi}(p)\right)^4 =\\
&(2\pi)^4 \delta^{(4)}(k_1+k_2+k_3+k_4) \,
(s^2+t^2+u^2) \, t_8^{a_1c_1 \dots a_4c_4} \zeta_1^{a_1} k_1^{c_1} \dots
\zeta_4^{a_4} k_4^{c_4} \ ,
\label{schm1}}
where the factor of $(2\pi)^4 \delta^{(4)}(k_1+k_2+k_3+k_4)$ comes
from $\int d^4 a$, and the remaining factor of
$(s^2+t^2+u^2) \, t_8^{a_1c_1 \dots a_4c_4} \zeta^{a_1} k_1^{c_1} \dots
\zeta^{a_4} k_4^{c_4}$ is the result of integrations over
$\int d^8{\cal M}$ and of a proper care over the index contractions
in \eqref{schm1}. An eight-rank tensor $t_8$ is an
appropriately symmetrized sum of products of four Kroneckers \cite{GG2}.

Finally, the $k$-instanton contribution to the 4-point scalar amplitude
on the 3-brane derived from the generating functional \eqref{finalrr} reads
\SP{&{\cal A}_{2 \to 2}^{\rm scalar} =\, {\rm const}\,{\alpha'}^4\,
\I_{\rm bulk}(k,1)\,
\delta^{(4)}(k_1+k_2+k_3+k_4) \\
&\times (s^2+t^2+u^2) (tu
\zeta_1\cdot\zeta_2 \zeta_3\cdot\zeta_4 +su
\zeta_1\cdot\zeta_4 \zeta_2\cdot\zeta_3+st
\zeta_1\cdot\zeta_3 \zeta_2\cdot\zeta_4 \big) \ ,
\label{amp4scs}
}
where const denotes some numerical factor independent of $k$,
and we have used $g_4^4/g_0^4 \propto {\alpha'}^4$. This amplitude has
precisely the same kinematic form as dictated by the Green-Gutperle
effective action \eqref{fourfmod}
\EQ{S_{\rm scalar}^{(4)\ \rm eff}=
{\rm const} \,{\alpha'}^4 \, \sum_{d|k} \frac{1}{d}\,
\int d^4x \sqrt{g^{(E)}}\,(\partial^2 \varphi)^4
\ ,
\label{seff4sc}}
hence we conclude that $\I_{\rm bulk}(k,1)=\sum_{d|k}d^{-1}.$


\begin{thebibliography}{99}
\small{
\bibitem{Witvarious}
E.~Witten,
``String theory dynamics in various dimensions,''
Nucl.\ Phys.\  {\bf B443} (1995) 85
[hep-th/9503124].

\bibitem{Schwarz} J.~H.~Schwarz,
``An $SL(2,{\mathbb Z})$ multiplet of type IIB superstrings,''
Phys.\ Lett.\  {\bf B360} (1995) 13
[hep-th/9508143].

\bibitem{OM} C. Montonen and D. Olive, {\em Phys. Lett.} {\bf 72B}
(1977) 117.

\bibitem{Rozali}
M.~Rozali,
``Matrix theory and U-duality in seven dimensions,''
Phys.\ Lett.\  {\bf B400} (1997) 260
[hep-th/9702136].

\bibitem{Berkooz}
M.~Berkooz, M.~Rozali and N.~Seiberg,
``On transverse fivebranes in M(atrix) theory on $T^{5}$,''
Phys.\ Lett.\  {\bf B408} (1997) 105
[hep-th/9704089].

\bibitem{manton}
N.~S.~Manton,
``A Remark On The Scattering Of BPS Monopoles,''
Phys.\ Lett.\  {\bf B110} (1982) 54.

\bibitem{Witsmall}
E.~Witten,
``Small Instantons in String Theory,''
Nucl.\ Phys.\  {\bf B460} (1996) 541
[hep-th/9511030].

\bibitem{Douglas1}
M.~R.~Douglas,
``Branes within branes,''
hep-th/9512077.

\bibitem{Douglas2}
M.~R.~Douglas,
``Gauge Fields and D-branes,''
J.\ Geom.\ Phys.\  {\bf 28} (1998) 255
[hep-th/9604198].

\bibitem{Diacon} D.~Diaconescu,
``D-branes, monopoles and Nahm equations,''
Nucl.\ Phys.\  {\bf B503} (1997) 220
[hep-th/9608163].

\bibitem{Impure}
A.~Kapustin and S.~Sethi,
``The Higgs branch of impurity theories,''
Adv.\ Theor.\ Math.\ Phys.\  {\bf 2} (1998) 571
[hep-th/9804027].

\bibitem{Windex}
E.~Witten, ``Constraints On Supersymmetry Breaking,''
Nucl.\ Phys.\  {\bf B202} (1982) 253.

\bibitem{Sethstern}
S.~Sethi and M.~Stern,
``D-brane bound states redux,''
Commun.\ Math.\ Phys.\  {\bf 194} (1998) 675
[hep-th/9705046].

\bibitem{Yi} P.~Yi,
``Witten index and threshold bound states of D-branes,''
Nucl.\ Phys.\  {\bf B505} (1997) 307
[hep-th/9704098].

\bibitem{Becker}
K.~Becker, M.~Becker and A.~Strominger,
``Five-branes, membranes and nonperturbative string theory,''
Nucl.\ Phys.\  {\bf B456} (1995) 130
[hep-th/9507158].

\bibitem{GG1}
M.~B.~Green and M.~Gutperle,
``D-particle bound states and the D-instanton measure,''
JHEP {\bf 9801} (1998) 005
[hep-th/9711107].

\bibitem{Sethgreen}
M.~B.~Green and S.~Sethi,
``Supersymmetry constraints on type IIB supergravity,''
Phys.\ Rev.\  {\bf D59} (1999) 046006
[hep-th/9808061].

\bibitem{Kiritsis}
E.~Kiritsis,
``Duality and instantons in string theory,''
hep-th/9906018.

\bibitem{PK}
B.~Pioline and E.~Kiritsis,
``U-duality and D-brane combinatorics,''
Phys.\ Lett.\  {\bf B418} (1998) 61
[hep-th/9710078].

\bibitem{part1} N. Dorey, ``Instantons, Compactification and S-duality
in ${\cal N}=4$ SUSY Yang-Mills Theory I'', hep-th/0010115.

\bibitem{part2} N. Dorey and A. Parnachev, ``Instantons, Compactification
and S-duality in ${\cal N}=4$ SUSY Yang-Mills II'', hep-th/0011202.

\bibitem{CV}
S.~Cecotti, P.~Fendley, K.~Intriligator and C.~Vafa,
``A New supersymmetric index,''
Nucl.\ Phys.\  {\bf B386} (1992) 405
[hep-th/9204102].

\bibitem{Nak}
H.~Nakajima,
``Resolution of moduli spaces of ideal instantons on ${\mathbb R}^4$,''
{\it  In *Sanda 1993, Topology, geometry and field theory* 129-136}.

\bibitem{Segal}
G.~Segal and A.~Selby,
``The cohomology of the space of magnetic monopoles,''
Commun.\ Math.\ Phys.\  {\bf 177} (1996) 775.

\bibitem{Witten1997}
E.~Witten,
``On the conformal field theory of the Higgs branch,''
JHEP {\bf 9707} (1997) 003
[hep-th/9707093].

\bibitem{Berkooz1999}
M.~Berkooz and H.~Verlinde,
``Matrix theory, AdS/CFT and Higgs-Coulomb equivalence,''
JHEP {\bf 9911} (1999) 037
[hep-th/9907100].

\bibitem{Aharony1999}
O.~Aharony and M.~Berkooz,
``IR dynamics of d = 2, N = (4,4) gauge theories and DLCQ of 'little
string theories,''
JHEP {\bf 9910} (1999) 030
[hep-th/9909101].

\bibitem{cal}
C.~Callias,
``Index Theorems On Open Spaces,''
Commun.\ Math.\ Phys.\  {\bf 62} (1978) 213.

\bibitem{MNS}
G.~Moore, N.~Nekrasov and S.~Shatashvili,
``D-particle bound states and generalized instantons,''
Commun.\ Math.\ Phys.\  {\bf 209} (2000) 77
[hep-th/9803265].

\bibitem{KNS}{W.~Krauth, H.~Nicolai and M.~Staudacher,
Phys. Lett. {\bf B431} (1998) 31, {\tt hep-th/9803117}\\
W.~Krauth and M.~Staudacher, Phys. Lett. {\bf B435} (1998) 350 , {\tt hep-th/9804199}}

\bibitem{Sen}
A.~Sen, ``Dyon - monopole bound states, selfdual harmonic forms
on the multi - monopole moduli space, and $SL(2,{\mathbb Z})$ invariance in
string theory,''
Phys.\ Lett.\  {\bf B329} (1994) 217
[hep-th/9402032].

\bibitem{gaunt}
J.~P.~Gauntlett,
``Low-energy dynamics of N=2 supersymmetric monopoles,''
Nucl.\ Phys.\  {\bf B411} (1994) 443
[hep-th/9305068].

\bibitem{blum}
J.~D.~Blum,
``Supersymmetric quantum mechanics of monopoles in N=4 Yang-Mills theory,''
Phys.\ Lett.\  {\bf B333} (1994) 92
[hep-th/9401133].

\bibitem{DKM97}
N.~Dorey, V.~V.~Khoze and M.~P.~Mattis,
``Multi-instantons, three-dimensional gauge theory, and the  Gauss-Bonnet-Chern
theorem,''
Nucl.\ Phys.\  {\bf B502} (1997) 94
[hep-th/9704197].

\bibitem{PSS} S.~Paban, S.~Sethi and M.~Stern,
``Summing up instantons in three-dimensional Yang-Mills theories,''
hep-th/9808119.

\bibitem{BFSS2}
T.~Banks, W.~Fischler, N.~Seiberg and L.~Susskind,
``Instantons, scale invariance and Lorentz invariance in matrix theory,''
Phys.\ Lett.\  {\bf B408} (1997) 111

\bibitem{AG}
L.~Alvarez-Gaume,
``Supersymmetry And The Atiyah-Singer Index Theorem,''
Commun.\ Math.\ Phys.\  {\bf 90} (1983) 161.

\bibitem{GH}
J.~P.~Gauntlett and J.~A.~Harvey,
``S duality and the dyon spectrum in N=2 superYang-Mills theory,''
Nucl.\ Phys.\  {\bf B463} (1996) 287
[hep-th/9508156].

\bibitem{EGH}
T.~Eguchi, P.~B.~Gilkey and A.~J.~Hanson,
``Gravitation, Gauge Theories And Differential Geometry,''
Phys.\ Rept.\  {\bf 66} (1980) 213.

\bibitem{ADHM} M.~F.~Atiyah, N.~J.~Hitchin, V.~G.~Drinfeld and Y.~I.~Manin,
``Construction of instantons,''
Phys.\ Lett.\ A {\bf 65} (1978) 185.

\bibitem{Nek}
N.~Nekrasov and A.~Schwarz,
``Instantons on
noncommutative ${\mathbb R}^4$ and $(2,0)$
superconformal  six  dimensional theory,''
Commun.\ Math.\ Phys.\  {\bf 198} (1998) 689
[hep-th/9802068].

\bibitem{LeeTong}
K.~Lee, D.~Tong and S.~Yi,
``The moduli space of two U(1) instantons
on noncommutative
${\mathbb R}^4$ and $R^{3}\times S^{1}$,''
hep-th/0008092.

\bibitem{partition}
N.~Dorey, T.~J.~Hollowood and V.~V.~Khoze,
JHEP {\bf 0103} (2001) 040
[hep-th/0011247].

\bibitem{Wittenphases}
E.~Witten, ``Phases of N = 2 theories in two dimensions,''
Nucl.\ Phys.\  {\bf B403} (1993) 159
[hep-th/9301042].

\bibitem{Aharony1998}
O.~Aharony, M.~Berkooz and N.~Seiberg,
``Light-cone description of (2,0) superconformal theories in six  dimensions,''
Adv.\ Theor.\ Math.\ Phys.\  {\bf 2} (1998) 119
[hep-th/9712117].

\bibitem{GG2}
M.~B.~Green and M.~Gutperle,
``D-instanton induced interactions on a D3-brane,''
hep-th/0002011.

\bibitem{Lee}
K.~Lee and P.~Yi,
``Quantum spectrum of instanton solitons in five dimensional
noncommutative U(N) theories,''
Phys.\ Rev.\ D {\bf 61} (2000) 125015
hep-th/9911186.

\bibitem{bstates} E.~Witten,
``Bound States Of Strings And p-Branes,''
Nucl.\ Phys.\  {\bf B460} (1996) 335
[hep-th/9510135].

\bibitem{KP}
E.~Kiritsis and B.~Pioline,
``On ${\cal R}^{4}$ threshold corrections in type IIB
string theory and (p,q) string  instantons,''
Nucl.\ Phys.\  {\bf B508} (1997) 509
[hep-th/9707018].

\bibitem{Kostov}
I.~K.~Kostov and P.~Vanhove,
``Matrix string partition functions,''
Phys.\ Lett.\  {\bf B444} (1998) 196
[hep-th/9809130].

\bibitem{MO3}
N. Dorey, T.J. Hollowood, V.V. Khoze, M.P. Mattis and
S. Vandoren,
Nucl. Phys. {\bf B552} (1999) 88
{\tt [hep-th/9901128]}.

\bibitem{Howe}
P.~Howe, K.~S.~Stelle and P.~K.~Townsend,
``Supercurrents,''
Nucl.\ Phys.\  {\bf B192}, 332 (1981).
}

\end{thebibliography}
\end{document}